\documentclass[journal=aesccq,manuscript=article]{achemso}
\setkeys{acs}{articletitle=true}
\setkeys{acs}{maxauthors=10}
\setkeys{acs}{etalmode=truncate}
\setkeys{acs}{doi=true}
\usepackage[T1]{fontenc} % Use modern font encodings
\usepackage[utf8]{inputenc}
\usepackage{lmodern}
\usepackage{subcaption}
\usepackage{float}
\usepackage{booktabs}
\usepackage{etoolbox}
\usepackage{textcomp}
\usepackage{multirow}
\usepackage{placeins}
\usepackage{geometry}
\usepackage{graphicx}
\usepackage{wrapfig}
\usepackage[numbers]{natbib}
\usepackage{rotating}
\usepackage{lineno}
\usepackage{xspace}
\usepackage{siunitx}
\DeclareSIUnit\Molar{\textsc{m}}
\usepackage{upgreek}
\usepackage{chemmacros}
\chemsetup{
	formula = mhchem,
	greek = chemgreek,
	modules = thermodynamics,
	modules = redox,
	modules = reactions
}

\newcommand{\chem}[1]{\ensuremath{\mathrm{#1}}}

\newcommand {\kcal} {kcal mol$^{-1}$\xspace}

\newcommand {\nacl} {\ce{[Na^+,Cl^-]} \xspace}
\newcommand {\na} {\ce{Na^+}}
\newcommand {\cl} {\ce{Cl^-}}
\newcommand {\supporting} {Supporting Information\xspace}
\newcommand {\etal} {et al.}

\author{Val\'erie Vallet}
\affiliation{Univ. Lille, CNRS, UMR 8523 - PhLAM - Physique des Lasers Atomes et Mol\'{e}cules, F-59000 Lille, France}
\author{Jonathan Coles}
\affiliation{Leibniz Supercomputing Centre of the Bavarian Academy of Sciences and Humanities (LRZ), Garching b. M\"unchen, Germany}
\author{Florent R\'eal}
\affiliation{Univ. Lille, CNRS, UMR 8523 - PhLAM - Physique des Lasers Atomes et Mol\'{e}cules, F-59000 Lille, France}
\author{C\'eline Houriez}
\affiliation{MINES ParisTech, PSL Research University,  CTP - Centre Thermodynamique des Proc\'ed\'es, 35 rue Saint-Honor\'e, 77300 Fontainebleau, France}
\author{Michel Masella}
\affiliation{Laboratoire Bio\'energ\'etique, M\'etalloprot\'eines et Stress, Service de Bio\'energ\'etique, Biologie Structurale et M\'ecanismes, Institut Joliot, CEA Saclay, F-91191 Gif sur Yvette Cedex, France}
\email{michel.masella@cea.fr}

\title{NaCl salts in finite aqueous environments at the fine particle marine aerosol scale.}

\keywords{} 

\begin{document}

\newpage

%\linenumbers

% Abstract

\begin{abstract}

We investigated isolated sodium/chloride aqueous droplets at the microscopic level, which comprise from about 5k to 1M water molecules and whose salt concentrations are 0.2$m$ (brackish water) and 0.6$m$ (sea water), by means of molecular dynamics simulations based on an \emph{ab initio}-based polarizable force field. The size of our largest droplets is at the submicron particle marine aerosol scale. From our simulations, we investigated  ion spatial distributions, ion aggregates (size, composition, lifetime and distribution), droplet surface potentials and the densities of the water vapor surrounding the droplets. Regarding ions, they form a weak electrostatic double layer extending from the droplet boundary to 2~nm within the droplet interior. Free {\na} and ion aggregates are more repelled from the boundary than free {\cl}. Most of the droplet properties depend on the droplet radius $R$ according to the standard formula  $A=A_\infty(1 -  2 \delta/R) $, where $A_\infty$ is the bulk magnitude of the quantity $A$ and $\delta$ is a length at most at the~nm scale. Regarding the water vapor densities they obey a Kelvin relation corresponding to a surface tension whose Tolman length is negative and at the 1~nm scale. That length is about one order of magnitude larger than for pure water droplets, however it is weak enough to support the reliability of a standard Kelvin term (based on planar interface surface tensions and water densities) and of the related K{\"o}lher equation to model sub-micron salty aerosols.

\end {abstract}

%\date{This manuscript was compiled on \today}

\newpage	

%Introduction

\section{Introduction}

Molecular scale phenomena at water/vapor interfaces are at the origin of strong modulation of many chemical reactions as compared to bulk~\cite{finlayson89,saykally13,ruiz20}. They are key factors to understand a large variety of important effects from enzyme catalysis~\cite{reis09,serrano18} to corrosion~\cite{stumm97} and they are thus of fundamental importance in atmospheric science. Besides their role in pollution effects~\cite{poschl15}, their role in atmospheric absorbing/scattering solar radiation as well as in cloud condensation~\cite{schill15} and ice formation~\cite{mccluskey17} explain their importance to understand and model climate change~\cite{rapport_giec21}.

Oceans are the main contributors of water aerosols (termed as Sea Spray Aerosols, SSAs) with radii that typically range from the 0.01 to the 10 $\mu$m scale\cite{saltzman09,herrmann15,quinn15,bertram18}. Fresh large SSAs (at the micron scale and above) are composed of unaltered sea water (a brine of sodium/chloride and magnesium sulfate salts that can also be enriched microorganisms \cite{aller04}) whereas sub-micron SSAs can be enriched by organic compounds (like fatty acids and sugars) compared to bulk sea water. Size distributions of SSAs depend on their location on earth. The diameters of the most abundant SSAs vary from 20 to 200~nm as reported by surface level measurements~\cite{farmer15}. In particular fine SSAs (at the 20 nm diameter scale) are reported to be the most abundant in marine boundary layers of the Southern Pacific \cite{bates98} and North Atlantic \cite{zhang14} oceans.

%The composition of SSAs evolves in the atmosphere by incorporating particles from different origins as well as by adsorbing water because of their overall large hygroscopicity~\cite{saltzman09}.  

As bulk systems, experimental and theoretical investigations of SSA properties, like the spatial distribution of their different components, double-layer formations, orientational and inter species cooperative effects, are highly challenging and still poorly documented, as recently reviewed by Bj\"ornehom and co-workers~\cite{bjornholm16}. Molecular dynamics, MD, numerical simulations are now widely used to investigate the microscopic properties of any kind of molecular systems. With the ongoing increase of the available computational resources, MD simulations of relatively large bulk molecular systems are now routinely performed, in particular to investigate microscopic phenomena at liquid water (planar) interfaces, see among others Refs.~[\citenum{mucha05,ishiyama07,dauria09,ottosson10,hammerich12,cummings13,neyt13,baer14,murdachaew16,olivieri18}]. 

The main difficulty that MD simulations face concerns the accuracy of the potential energy terms (the force field for classical MD) that are used to model microscopic forces. The parameters of most of the available force fields, in particular those used to simulate salty aqueous systems, are adjusted to reproduce experimental data, like osmotic pressures~\cite{luo13,soniat16}. Besides the force-field sophistication (like accounting explicitly for microscopic polarization), the latter strategy to build force fields gives rise questions about their transferability, \emph{i.e.} their ability to accurately describe molecular systems in physical states that were not considered to adjust them. To tackle these difficulties, new classes of force fields, termed \emph{ab initio}-based force fields and whose parameters are adjusted only by considering \emph{ab initio} quantum chemistry data regarding small molecular systems~\cite{xu18}, are inferred to be potentially more accurate and to allow investigations of molecular systems for which no experimental data are available.  

As all the studies mentioned above, most of the available MD simulations of salty aqueous solution/gas phase boundaries  regard planar interfaces simulated using periodic boundary conditions. The size of the simulation boxes are overall small: they typically comprise from \num{2000} to \num{10000} water molecules and ions. Nevertheless they are shown to provide reliable data regarding the properties of planar interfaces and thus, as usually inferred, of spherical interfaces corresponding to large enough finite size molecular systems, at the 10$^{6}$ molecules scale and above \cite{granasy98}. Assuming the water density within SSAs to match the liquid water one (33.3 molecules~nm$^{-3}$ at ambient conditions), the radius of a SSA comprising 10$^{6}$ molecules is about 20~nm. From the size distributions of SSAs mentioned above, simulations of planar interfaces are thus well suited to investigate the properties of a sizable part of the SSAs but not of the finest ones that are far from being negligible and that may even be the most abundant at ocean surface levels. 

Moreover there are still controversies about the reliability of all the laws related to the Gibbs classical thermodynamic framework of interfaces (\emph{i.e.} interfaces that correspond to large surfaces separated by thin boundary layers) to model sub-micron SSAs, as recently discussed in Refs. \cite{liu16,montero20,kashchiev20,miguel21} (see also the references cited therein). Among these laws, we may cite the Tolman relation to compute the surface tension of spherical finite droplets \cite{tolman49}, a key parameter of the K\"ohler equation to quantify the ability of a SSA to evolve towards a cloud condensation nucleus  \cite{koehler36,farmer15}. 

Regarding liquid-like finite size systems and besides Functional Density approaches \cite{mcgraw96, kashchiev03}, only one component homogenous systems, like pure aqueous small droplets \cite{joswiak13,factorovich14,leong18} (up to the 10$^4$ molecules scale) and pure Lennard-Jones fluids \cite{moody03,wilhelmsen15} (up to the 10$^6$ molecules scale), have been investigated by means of microscopic simulations to address the above issue regarding the reliability of the laws related to the Gibbs framework. Interestingly and in agreement with recent experiments regarding $n$-propane \cite{zhong18}, the latter simulation studies concluded to the reliability of such laws, even to model water aggregates comprising only a few hundred molecules \cite{factorovich14}.

Our aim here is to discuss the ability of the following simple equation to model any property $A$ of quasi-spherical sub-micron SSAs whose radius is $R$ :

\begin{equation} \label{eqn:property}
A(R)  \approx  A_\infty (1 - \dfrac{2 \delta_A}{R}) +O(\dfrac{1}{R^2}),
\end{equation}
here $A(R)$ and $A_\infty$ are the numerical quantities measuring the magnitude of the property $A$ for spherical and planar interfaces, respectively, and $\delta_A$ is a length. Note that,  for large enough radius $R$, we may readily rewrite the Tolman relation to match the above equation. 

To this end we present MD simulations at ambient temperature of ideal fine SSAs (a brine comprising only NaCl salt) whose molecular size varies from about 5.10$^3$ (5k) up to 10$^6$ (1M) water molecules. The simulations were performed by means of the latest version of our \emph{ab initio}-based polarizable force field~\cite{real13,real16,real19,houriez19}. We focus our discussions on properties like surface potential, ion distribution and ion clustering. To  assess the accuracy of our simulation protocol, we systematically extrapolate bulk quantities from our salty SSA data and we compare them  to available experimental and simulation results regarding planar interfaces of salty solutions. Lastly, we also discuss water densities in the vapor phase surrounding our SSAs, which allows us to discuss the reliability of the Kelvin relation, a key component of the K\"ohler equation, for salty droplets at the sub-micron scale.

Because of the size of our largest droplets (at the 1M molecules scale) and to further assess our simulation protocol, we not only consider systems corresponding to a salt concentration of 0.6$m$ (as sea water) but also of 0.2$m$. Besides to investigate properties of SSAs originating from brackish waters, simulations of large 0.2$m$ systems will provide new insights regarding weakly concentrated NaCl solutions (like ion clustering) that are usually not investigated using standard MD protocols that focused on molar and above salt solutions (see the studies mentioned above). Note that 0.6$m$ salty droplets correspond to fresh marine SSAs, whereas aged SSAs evolve towards more concentrated salt systems in the atmosphere \cite{bertram18}.

Lastly, as discussed by Debiec~{\etal}\cite{debiec14} for instance, the Potential of Mean Force, PMF, corresponding to an ion pair dissolved in neat water is an important parameter to assess the reliability of  force fields devoted to investigate salt hydration. We will thus also discuss the PMF in bulk water of a single \na/\cl pair  computed from our force fields at ambient conditions by comparing its main features to the predictions of well-accepted theoretical approaches. 

\section{Theoretical details} \label{sec:force_field}

\subsection{Modeling ion/water interactions}

To model hydrated sodium/chloride salts, we consider our own  \emph{all-atom} polarizable  force fields based on a rigid water representation and for which the total potential energy $U$ is decomposed in five terms 

\begin{equation} \label{eqn:total-energy}
U = U^{rep} + U^{qq'} + U^{pol} + U^{coop} + U^{sc}.
\end{equation}
The first two terms correspond to short-range repulsion and Coulombic interactions 
\begin{gather*}
U^{rep} = \sum_{i=1}^{N}\sum_{j>i}^{N^*} A_{ij}\exp{(-B_{ij}r_{ij})},\\
U^{qq'} = \sum_{i=1}^{N}\sum_{j>i}^{N^*} \frac{q_iq_j}{4\pi\varepsilon_0r_{ij}}.
\end{gather*}

Here $r_{ij}$ is the distance between the pair of atoms ($i$,$j$), $q_i$ are the static charges located on the atomic center $i$, and $A_{ij}$ and $B_{ij}$ are adjustable parameters. $N$ is the total number of atoms within the molecular system and the superscript $^*$ indicates that the corresponding sum does not account analytically for water intramolecular interactions.  The polarization term $U^{pol} $ is based on an induced dipole moment approach. The induced dipole moments $ \{\mathbf{p}_i\}_{1\leq i\leq N_\mu}$, one per non-hydrogen atom/ion, obey
\begin{equation}
\label{eqn:induceddipole}
\mathbf{p}_i = \alpha_i\cdot\left(\mathbf{E}_i^q + \sum_{j=1}^{N^*_\mu}{\mathbf{T}_{ij}\cdot\mathbf{p}_j}\right).
\end{equation}

The static charge electric field $\mathbf{E}_i^q$ acting on a polarizable atom $i$ is generated by the above charges $q_i$. $\alpha_i$ is the isotropic polarizability of water and ions, and $\mathbf{T}_{ij}$ is the dipolar interaction tensor. Both $\mathbf{E}_i^q$ and  $\mathbf{T}_{ij}$ include short-range Thole-like damping functions~\cite{thole81}. The set of Eqs.~(\ref{eqn:induceddipole}) is iteratively solved and the resulting polarization energy is 
\begin{equation} \label{eqn:pol}
U^{pol} = \frac{1}{2}\sum_{i=1}^{N_\mu} \dfrac{\mathbf{p}^2_i}{\alpha_i}-\sum_{i=1}^{N_\mu} \mathbf{p}_i \cdot \mathbf{E}_i^{q} - \dfrac{1}{2} \sum_{i=1}^{N_\mu}\sum_{j=1}^{N^*_\mu} \mathbf{p}_i \mathbf{T}_{ij} \mathbf{p}_j.
\end{equation}

To accurately model cooperative effects occurring in water hydrogen bond networks and to account for particular electronic density reorganization effects between the water molecules and halide anions, our force fields include a set of short-range cooperative energy terms denoted  $U^{coop}$~\cite{trumm12,real13,houriez15} 
\begin{equation}
U^{coop} = \sum D_e^* f(r)g(\{\theta\}).
\end{equation}
The sum runs on water/water and water/chloride hydrogen bonds, HBs, $r$ is the HB length and $f$ is a Gaussian function. $\{\theta\}$ is a set of angles allowing one to model the directionality of HBs and the corresponding function $g$ is a product of Gaussian functions of these angles. The originality of our $U^{coop}$ term arises from $D_e^*$ that is not a static parameter: its magnitude is modulated by the chemical environment of the entities involved of a specific HB (precisely by the local density of water molecules at the vicinity of those entities)~\cite{trumm12,real13,houriez15}. 

Our force-field  parameters to model the hydration of anions and cations are assigned to reproduce quantum \emph{ab initio} data regarding relatively small \ce{[X^-,(H_2O)_n]} clusters ($n \leq 6$). For that purpose we only consider quantum data from the MP2 level of theory extrapolated to the Complete Basis Set, CBS, limit. Lastly, we further improve the reliability of our force fields to accurately model large anion/water clusters ($n > 6$ and for which all the water molecules lie in the anion first hydration shell) and alkali cation/water clusters presenting a partial or complete second hydration shell by means of short range many-body functions $U^{sc}$, whose analytical form is close to the $U^{coop}$~\cite{real19,houriez19} (see also the details provided as \supporting). For the present purpose, we used the $U^{sc}$ functions (and their parameters) allowing one to model the hydration of \ce{Na^+} and \ce{Cl^-} from our earlier studies~\cite{real19,houriez19}.
 
Regarding the full series of halides (from \chem{F^-} to \chem{At^-}), we showed the ability of our force fields to reproduce their structural (like anion/water first coordination shell), energetic (hydration enthalpy and Gibbs free energy) and temporal (ion coefficient of diffusion) properties in neat water at ambient conditions \cite{real16,real19,houriez19}.  In particular our force fields do not yield over polarization effect regarding halides~\cite{trumm12} contrary to the present version of the AMOEBA approach~\cite{rogers10}, another promising class of  polarizable force fields. We have also shown the ability of our force field to model structural and energetic properties regarding cations \chem{Li^+}, \chem{K^+} and \chem{NH_4^+} and of its methylated forms \cite{houriez14,houriez19}. Regarding  \chem{Na^+}, we provide new data regarding to its hydration structural properties as computed from our droplet simulations in \supporting. These data are in line with experiment \cite{galib17}.
 
\subsection{The force field to model \ce{Na^+/Cl^-} interactions}

Inter ionic \ce{Na^+/Cl^-} interactions are modeled by considering only the three energy terms $U^{rep}$, $U^{qq'}$, $U^{pol}$ detailed above. We assigned the repulsion $A_{{\na},{\cl}}$ and $B_{{\na},{\cl}}$ and polarization damping parameters to best reproduce quantum MP2/CBS data corresponding to inter ionic distances ranging from 0.2 to 0.7~nm, namely the ion pair interaction energies and the ion pair total dipole moments. Here CBS data are extrapolated from quantum computations at the MP2 level using the aug-cc-pVXZ (X=T,Q and 5) basis sets and performed using the MOLPRO package of programs~\cite{MOLPRO-WIREs,molpro}.

We compare in Figure~S3 of \supporting the interaction energies of the neutral dimer \ce{[Na^+,Cl^-]} and of the hetero charged trimers \ce{[(Na^+)_2,Cl^-]} and \ce{[Na^+,(Cl^-)_2]} from our force field and quantum computations. A 0.5 \% agreement is achieved. That represents an absolute error weaker than 0.5 (dimer) and 1 (trimer) \kcal regardless of the ion distance separation. Regarding the \ce{[Na^+,Cl^-]} dipole, a similar force-field/quantum agreement is also achieved on average. Even if that does not prejudge of the force-field quality in modeling larger \ce{[Na^+,Cl^-]} ionic clusters, the latter results support the overall accuracy of our modeling approach.

\subsection {Molecular Dynamics details}

MD simulations are performed at 300~K in the NVT ensemble using the code POLARIS(MD)~\cite{polaris}.  All the covalent \ce{O-H} bonds and \ce{H-O-H} angles are constrained to their force-field reference values by means of the iterative RATTLE procedure (the convergence criterion is set to 10$^{-7}$~nm). Induced dipole moments are iteratively solved until the mean difference in these dipoles between two successive iterations is less than 10$^{-6}$ Debye. The Newtonian equations of motion are solved using a multiple-time-steps algorithm~\cite{masella06}. From preparatory runs, a reasonable total energy drift is observed along MD trajectories using a time step of 2~fs for intermolecular short-range interactions (corresponding to a cutoff distance of 0.8~nm) and a time step of 6~fs to account for the remaining intermolecular long-range interactions (see Figure~S4 of \supporting). These time steps are systematically used for our study. Lastly, the system temperature is monitored using the General Gaussian Thermostats~\cite{liu00}. 

We compute the Coulombic and polarization energies using a Fast Multipole Method, FMM, scheme devoted to induced dipole-based polarizable force fields~\cite{coles15}. Here we set the FMM opening angle and the multipole expansion order to 0.7 and 7, respectively. As our FMM scheme does not conserve the system total angular momentum, we periodically rescale it so that the kinetic temperature corresponding to the global system rotational degrees of freedom is smaller than 1~K.

To model droplets in equilibrium with a vapor phase, we embedded them in a spherical cavity whose radius $R_\mathrm{cavity}$ corresponds to half the largest inter atomic distance within the simulation starting structures to which the distance $d =1.2$~nm is added. The cavity center is set to droplet Center Of Mass, COM. As a water molecule crosses the cavity boundaries it undergoes the harmonic potential $U^\mathrm{cavity} = k_\mathrm{cavity}(r-R_\mathrm{cavity})^2$ for $r \geq R_\mathrm{cavity}$ (and $U^\mathrm{cavity} \equiv 0$ otherwise). $r$ is the distance between COM and the water oxygen, and $k_\mathrm{cavity}$ is set to 100 \kcal~{\AA}$^{-2}$. The starting structures correspond to water molecules and ions set on the nodes of a cubic grid (the distance among the nodes is 0.35 nm). During the early simulation stages, these structures evolve towards spherical droplets whose surface is located from 2 (5-10k) up to 5 (1M)~nm from the cavity boundary.

We computed the Potential of Mean Force, PMF, using a standard Umbrella Sampling approach of a single \ce{[Na^+,Cl^-]} pair dissolved a cubic box comprising \num{1000} water molecules using periodic boundary conditions.  Coulombic and polarization interactions are computed according to a Smooth Particle Mesh Ewald scheme~\cite{toukmaji00} (the direct term energy cut off and the grid dimension are set to 1.2 and 0.1, and the interpolation order is 8). The details of the Umbrella Sampling approach may be found in our earlier studies (see Ref.~\citenum{houriez17_2} for instance). 

The MD trajectories are sampled each 10 ps once an initial relaxation phase of 10~ns is achieved.  When not detailed, the error regarding the averages discussed below are computed from a basic block analysis scheme. For 10k and 100k droplets, we compute averaged quantities over 10~ns MD segments and we assume the error to be the standard deviation corresponding to those averages.  For 1M droplets, we perform the analysis on 5~ns blocks. 

\subsection{The hydrated NaCl salts simulated} \label{sec:simulated_systems}

\begin{figure}[H]
\includegraphics[scale=1.0]{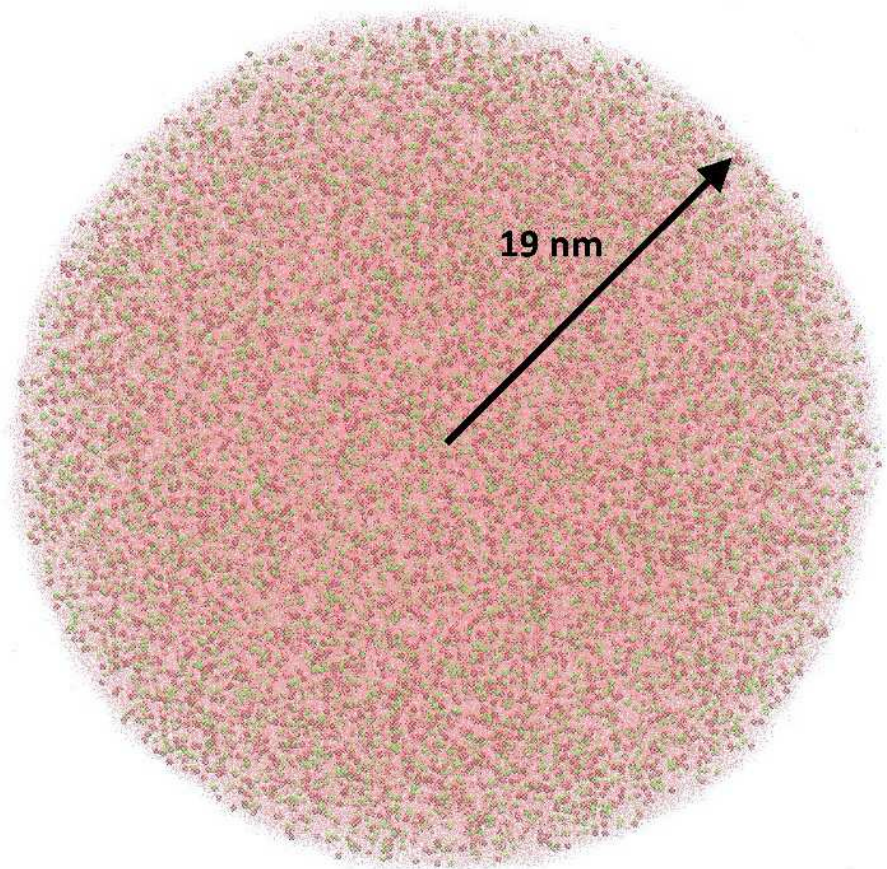}
 \caption{Relaxed structure of the 1M droplet corresponding to a salt concentration of 0.6$m$ (and comprising \num{10000} \nacl pairs). Water molecules are shown in red in a transparency mode, whereas green and red spheres correspond to ions.}  \label{fig:structure_1M}
 \end{figure}

We simulated two sets of NaCl/water droplets corresponding to salt concentrations of 0.2 and 0.6$m$ and comprising from about 5k to 1M water molecules at the 200~ns scale (5k-10k), 100~ns scale (20k-100k) and 30~ns (1M). The largest 1M droplets comprise 4k and 10k \ce{[Na^+,Cl^-]} pairs, respectively, see also a simulation snapshot of the 1M 0.6$m$ water droplet shown in Figure~\ref{fig:structure_1M}.

We also simulated smaller salty 0.2 and 0.6$m$ droplets at the 1k and 2k water molecules scale up to 500~ns. Our simulations show 5k droplets to be the smallest systems whose structural properties are close to those of larger systems, whereas the properties of 1k/2k 0.2$m$ droplets differ noticeably from larger ones and 1k/2k 0.6$m$ droplets are unstable. We will below briefly discuss the data corresponding to those 1k/2k systems.

% Results

\section{Results and discussion}

\subsection{NaCl ion pairing in aqueous phase}

We discuss here the {\nacl} ion pairing in liquid water from the main features of the PMF of a single dissolved {\nacl} pair in a periodically replicated box comprising about \num{1000} water molecules. That PMF has been already intensively investigated in particular by means of quantum Car-Parinello Molecular Dynamics, CPMD, approach~\cite{timko10}, and by means of polarizable force fields based on Drude oscillators and whose parameters were refined to reproduce experimental osmotic pressure data~\cite{luo13}. 

As for most ion pairs~\cite{pliego20} and as predicted by most of the simulations reported to date for the \nacl pair (see among others Refs.~[\citenum{fennel09,kelley15}]), both the above simulation approaches show that PMF to present a first minimum (termed Contact Ion Pair, CIP) at a short inter ionic distance $R_i$ (about 0.27~nm), then a first maximum (denoted Transition State, TS) at $R_i \approx 0.38$~nm and a second minimum (Solvent Separated Ion Pair, SSIP) at $R_i \approx 0.55$~nm. By setting to zero these PMFs at an inter-ionic distance of 0.6~nm, CPMD predicts the CIP and SSIP energy depths to be about -0.5 $\pm$ 0.4~\kcal, whereas the most accurate Drude-based force field predicts these depths to be 0.1 and -0.5 $\pm$ 0.1~\kcal, respectively. Lastly, the energy barrier height of TS is about +1.4~\kcal for both approaches. However regarding CPMD results, a recent study shows the overall strong sensitivity of that PMF to the available DFT functionals~\cite{wills21}.

\begin{figure}[H]
\includegraphics[scale=.7]{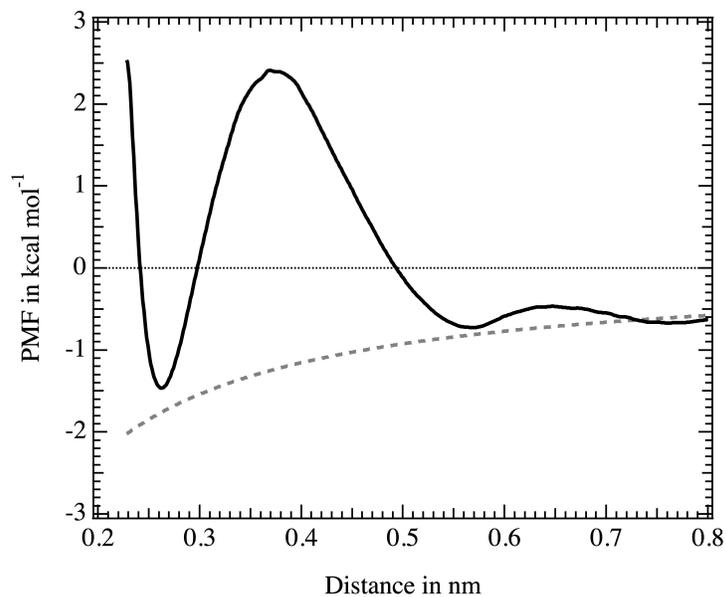}
 \caption{{\nacl} ion pair PMFs in neat liquid water at ambient conditions as computed from our force fields (black line). Dashed grey line, the expected Coulombic PMF corresponding to two point charges of opposite sign in bulk water at ambient conditions. Our simulated PMF is shifted to meet the Coulombic PMF value at 0.8~nm.}  \label{fig:pmf}
 \end{figure}

In Figure \ref{fig:pmf} we plot  the {\nacl} PMF in liquid water computed from our polarizable force-field approach. Its main features agree with the standard locations of the CIP, TS and SSIP states. Moreover the depths of the CIP and SSIP minima (by shifting to zero the PMF value at $R_i  = 0.6$~nm) are in a reasonable agreement with their CPMD and Drude-based counterparts: about -0.9 (CIP) and -0.2 (SSIP) \kcal. The most noticeable difference between our approach and earlier CPMD and Drude-based PMF data concerns the TS energy height: our approach predicts it to be larger by 1.6 \kcal than the later ones. In all our PMF agrees with those computed from CPMD and Drude-based force-field simulations, even we if we may expect our force field to slightly more favor {\nacl} ion pairing in liquid water than the latter two approaches (because of the larger energy depth that our force field predicts for CIP).  

Our polarizable force field is built to reproduce high level \emph{ab initio} MP2/CBS data regarding small water and ion/water clusters in the gas phase. Hence together with the CPMD scheme~\cite{timko10} and Drude-based force fields~\cite{luo13} refined to reproduce experimental data, all these theoretical methods predict weak energy depths (at the k$_\mathrm{B}$T scale and weaker) regarding the {\nacl} CIP and SSIP minima in liquid water. 

\subsection{Droplet interfaces and ion distributions}

\begin{figure}[H]
\includegraphics[scale=.8]{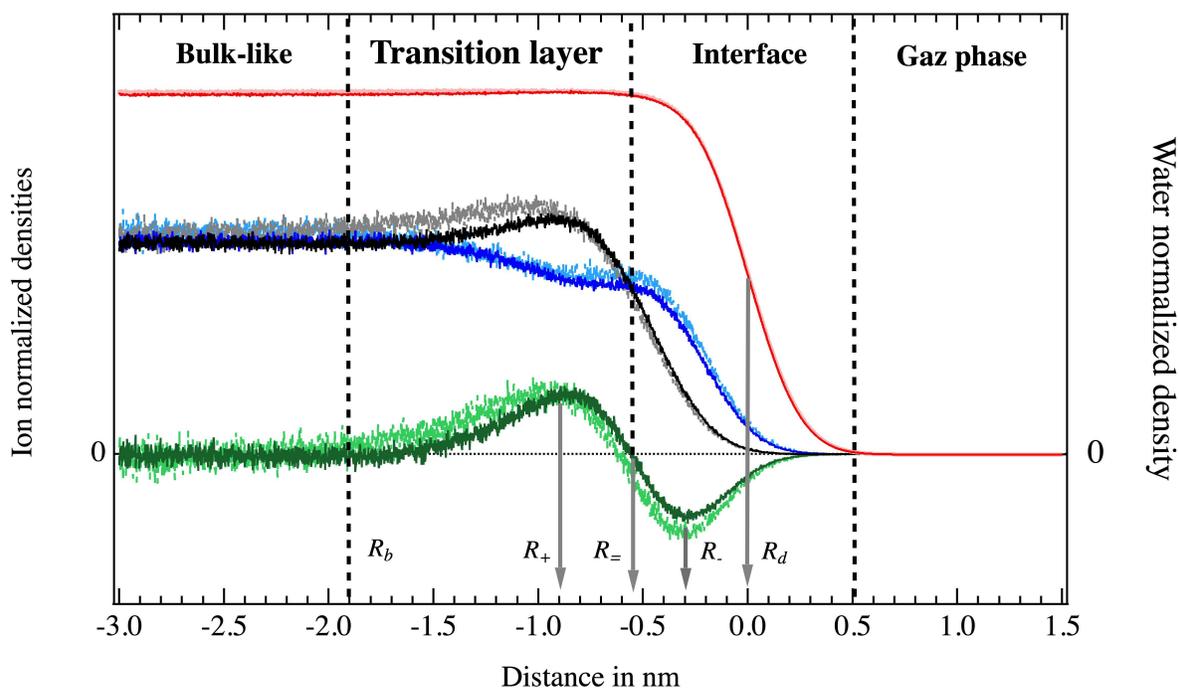}
 \caption{Ion (left axis) and water (right axis) normalized densities as a function of the distance from the mean droplet radius $R_d$ for 100k droplets. Negative distances correspond to the droplet interior. Water, \ce{Na^+} and \ce{Cl^-} densities are plotted in red, black and blue lines, respectively. 0.2$m$ data are plotted in lighter colors than 0.6$m$ ones. Green lines: differences $\Delta \rho^i(r)$ in the \ce{Na^+} and \ce{Cl^-} densities. The vertical grey arrows are located at the minimum, zero and maximum values of the $\Delta \rho^i(r)$ functions and at the mean droplet radius. The vertical black dashed lines delimit the four main domains from the droplet core to the gas phase. The profiles for 5k and 1M droplets are fully in line with the present 100k ones (see Supporting Material).}
  \label{fig:densities}
 \end{figure}

The water oxygen and ion mean density functions $\rho(r)$ and $\rho^i(r)$ ($i=$ {\na}, {\cl}) computed at a distance $r$ from the droplet COM are close regardless of the system size and the salt concentration. We plot these functions for 100k droplets in Figure \ref{fig:densities} whereas the 5k, 10k, 20k and 1M functions are plotted in Figure~S5 of \supporting. The water densities  $\rho(r)$ exhibit the usual sigmoidal  behavior

\begin{equation}
\rho(r) = \dfrac{\rho_d + \rho_g}{2} +  \dfrac{\rho_d - \rho_g}{2} \left[ \tanh \left( (r- R_d) / \xi  \right)  \right].
\end{equation}
Here, $\rho_d$ and $\rho_g$ are the mean water densities within the droplet core and the surrounding gas phase. $\xi$ is the thickness of the droplet/gas phase interface, respectively. $R_d$ is the radius of the equimolar dividing surface. To prevent the spurious effects arising from poor statistics at the droplet center,  we first adjusted the latter parameters to best reproduce $\rho(r)$ data within  -1 to +1~nm from the ideal droplet radius computed from the mean density of neat water at ambient conditions. Then we refine the $\rho_d$ values by averaging $\rho(r)$ within -2 to -1~nm from $R_d$. The resulting parameters are summarized in Table \ref{tab:droplet_data}.   

Regardless of the droplet size and the salt concentration, the thickness $\xi$ increases from 0.19 (5k) to 0.25~nm (1M) and the radius $R_d$ varies from 3.2 (5k) to 19.2~nm (1M). The radii of 1M droplets are in line with the lower-bound size of the most abundant sub-micron SSAs~\cite{herrmann15,bertram18}. Below we measure the distances $r$ from the radii $R_d$. Negative $r$ values indicate the depth into droplet interiors. All the water densities $\rho_d$ at the droplet core are close to the neat water one. They are  linearly but weakly correlated to the droplet curvature according to relation (\ref{eqn:property}): the magnitude of the corresponding length $\delta_\rho$ is about -0.02 (0.6$m$) and -0.1 (0.2$m$)~nm. Curvature effects are thus responsible of stronger pressures within the droplet cores as expected \cite{leong18}.

The mean ionic densities $\rho^i(r)$ differ at the vicinity of the droplet boundaries for our two kinds of ions (see Fig.~\ref{fig:densities}). {\na} are repelled from the droplet surface and they tend to slightly accumulate at a distance included between -1.5 and -0.5~nm from $R_d$. {\cl} are repelled in a lesser extent from the boundaries, however, their density is lower than within droplet cores. The densities of the two ions are half those within the droplet cores at 0.3 ({\cl}) and 0.4 ({\na})~nm before the droplet boundary (regardless of the droplet size and salt concentration) in agreement with earlier experimental-based estimates (that range from 0.31 up to 0.39~nm)~\cite{aveyard76}.

We quantify the difference in the ion behavior at the droplet boundary vicinity from the difference $\Delta \rho^i(r)$ in the {\na} and {\cl} densities $\rho^i(r)$. From the plots of Figure \ref{fig:densities}, the functions $\Delta \rho^i(r)$ all present the same profiles: two symmetric peaks at $R_- = -0.3$~nm (peak corresponding to the {\cl} predominance domain) and at $R_+=- 0.9$~nm ({\na} predominance domain) from $R_d$. Within the droplet core and starting at $R_+$, these functions converge then slowly to zero (the convergence is not achieved before $r=-1.9$~nm from $R_d$). These data show the existence of a weak electric double layer with a separation of about 0.6~nm, and starting at 0.3~nm from the droplet surfaces. Regardless of the salt concentration, the differences among the functions $\rho^i(r)$ for a given ion kind are weak from 10k to 1M droplets. Regarding 5k droplets, their functions $\rho^i(r)$ overall agree with those of larger systems. However in the particular case of the 0.6$m$ 5k droplet, we note the function $\rho^{Na}(r)$ to be flatter compared to larger systems within the whole transition layer domain (see  Figure~S5 of \supporting).

$\Delta \rho^i(r)$ functions lead us to define four main domains valid for all our droplets: the droplet core, that extends from the droplet COM up to -1.9~nm from the droplet radius $R_d$;  a transition layer, extending from -1.9 to -0.6~nm from $R_d$ in which \ce{Na^+} are slightly dominating; the droplet interface, extending from -0.6 to +0.5~nm  from $R_d$ in which \ce{Cl^-} are dominating in its first half, and above, the gas-phase domain. 
 
 The behavior of our ions at the droplet boundary is in line with most of the available data reported from polarizable MD simulations regarding {\nacl} aqueous solutions at a larger but still moderate concentration 1$m$~\cite{ishiyama07,cummings13,neyt13}, even if overall large differences in the  {\cl} behavior at the close vicinity of the droplet surface exist among these studies. For instance, the $\rho^i(r)$ density for {\cl} can be lower  within the full droplet interface domain than within the bulk region~\cite{cummings13} like in our simulations, or to be marked by a noticeable peak in that domain, peak whose height is larger than the {\cl} core density~\cite{ishiyama07,neyt13}. We also note a good agreement between our results and earlier non-polarizable simulation data~\cite{dauria09,ottosson10,olivieri18} that were shown  to reproduce  accurately photoelectron spectroscopy measurements regarding {\nacl} aqueous solutions within the salt concentration range 0.7 - 2.0$m$~\cite{ottosson10,olivieri18}. Note also most of the reported experiments, usually performed at higher salt concentrations ($\geq 2m$) to conclude to a weaker repulsion of {\cl} from the droplet boundary than {\na}~\cite{knipping00,tian11,piatkowski14}, even if there are controversies regarding the magnitude of that differential behavior~\cite{bjornholm16}.
  
Lastly, regarding 0.2$m$ droplets at the 1k-2k molecular scale, they are stable along multiple independent simulations performed at the 500~ns scale. Their radii $R_d$ are 1.9 and 2.4~nm, respectively, and the ion distributions at the droplet surface vicinity are in line with those of larger systems. However there is a noticeable difference within the droplet core:  {\na} cations are there more abundant than {\cl} anions by 15\% (see Figure~S6 of \supporting). Within those small salty droplets there are thus only two domains, the first close to the interface in which \cl dominates and a second interior domain where \na dominates. The ion density inhomogeneity within the droplet core is a source of potential instability arising from instantaneous inter ionic repulsive Coulombic effects. To our opinion that explains why 1k-2k 0.6$m$ droplets are unstable: they systematically and rapidly split (within 50~ns) in subsystems along all the multiple independent simulations that we performed.   
  
\subsection {Ion surface excess}

Assuming the gas phase density $\rho_\mathrm{g}^i$ of a component $i$ to be negligible compared to its droplet core density $\rho_\mathrm{d}^i$, the surface excess $\mathrm{\Gamma_i }$ of that component can be written as~\cite{ho03}
\begin{equation} \label{eqn:excess}
\Gamma_i = \dfrac{N_i}{4\pi R_e^2} - \dfrac{R_e \rho_\mathrm{d}^i} {3}.
\end{equation}
Here, $N_i$ is the total number of entities $i$. We set the radius $R_e$  to zero the water surface excess. From the water $\rho_d$ values summarized in Table \ref{tab:droplet_data}, we get $R_e$ values that agree with the $R_d$ ones within $\pm$ 0.03~nm.

Regarding ions, their droplet core densities $\rho_d^i$ are their mean densities  within a spherical shell $\mathbf{S}_d$ extending from $R_d - 3$ to $R_d - 2$~nm  (see Table \ref{tab:droplet_data}). Regardless of the salt concentration, these densities are linearly correlated to the droplet curvature, \emph{i.e.} they obey the standard relation (\ref{eqn:property}) (see Figure~S8 of \supporting). The length $\delta_\rho^i$ is positive and at the 1~nm scale for both our ions. The $\rho_d^i$ densities thus decrease as the droplet size increases. We interpret that behavior as arising from the ion depleted domain at the close vicinity of the droplet boundary and whose extension ($r_\mathrm{exc} \approx 0.3$~nm) does not depend on the droplet size. Our ions are thus mainly located within a spherical subdomain of our droplets whose radius is $R_d - r_\mathrm{exc}$. That explains the "accessible" volume for our ions (corresponding to the latter spherical subdomains) to be proportionally smaller for small droplets than for larger ones. For instance, the ion "accessible" volume is about 40\% smaller than the total droplet volume (as defined from $R_d$) for 5k droplets, and only 5\% smaller for 1M droplets. 

In Table \ref{tab:droplet_data} we summarized the ion surface excess $\Gamma_i$ computed from Equation (\ref{eqn:excess}) for our droplets. The upper bound of the uncertainties corresponding to the densities $\rho_d^i$  (as computed from a basic block analysis where the domain $\mathbf{S}_d$ is decomposed in ten subdomains of thickness 0.1~nm) amounts to 1\%. We use that value together with the mean difference value $R_e - R_d = 0.03$~nm (that we assume to be the uncertainty on the droplet radius $R_e$) to estimate the uncertainties on $\Gamma_i$, which are relatively large, about 20 \% of the $\Gamma_i$ values summarized in Table \ref{tab:droplet_data}.

\begin{figure}[H]
\includegraphics[scale=.8]{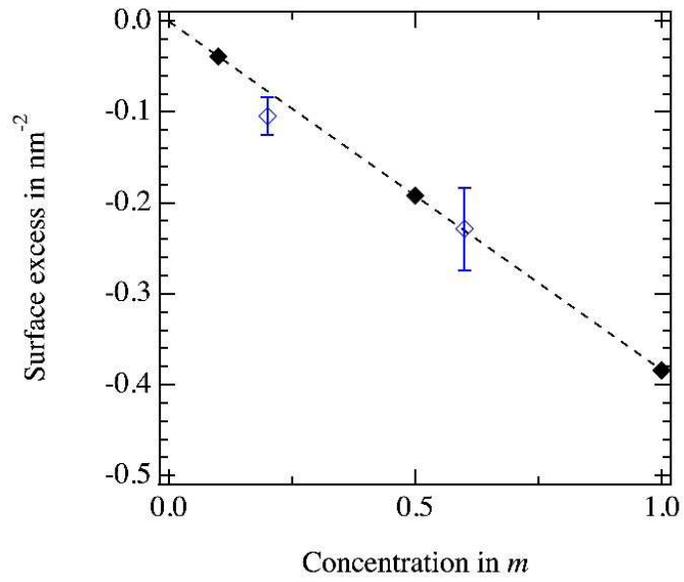}
 \caption{Up: total ion surface excess values $\Gamma_\mathrm{tot}$ as a function of the salt concentration. Blue and black: 0.2 and 0.6$m$ data, respectively; dashed lines: the corresponding  linear regression fits. Down: converged bulk values for $\Gamma_\mathrm{tot}$ (blue symbols)  compared to the experimental-based estimates \cite{shah13} (black symbols, the dashed lines correspond to the linear regression fit of the experimental values).}
  \label{fig:surface_excess}
 \end{figure}

For 10 k to 1M droplets, the sum $\Gamma_\mathrm{tot} = \Gamma_{\ce{Na^+}} + \Gamma_{\ce{Cl^-}}$  is a linear increasing function of $R_e^{-1}$, see Figure \ref{fig:surface_excess}, and their extrapolated bulk limit values are  -0.104 $\pm$ 0.020 and -0.228 $\pm$ 0.056~nm$^{-2}$ for 0.2 and 0.6$m$ systems, respectively. These bulk values are in a good agreement (within the error bars) with recent experimental-based estimates~\cite{shah13}, see Figure \ref{fig:surface_excess}, and a priori  in a better agreement than values from non-polarizable force fields~\cite{underwood18}. Regarding 5k droplets, the 0.2$m$ $\Gamma_\mathrm{tot}$ value obeys the above linear relation, whereas the 0.6$m$ value does not align with larger system data. That suggests a dependence of $\Gamma_\mathrm{tot}$ on higher order $R_e^{-n}$ terms for small salty droplets corresponding to molar and larger salt concentrations. However for typical sub-micron SSAs, the standard relation (\ref{eqn:property}) is valid for the quantity $\Gamma_\mathrm{tot}$.

\subsection {Salt clusters}

At a time $t_0$, we define a salt cluster as a set of \na {} and \cl {} ions that are all located at distance shorter than $d_\mathrm{cluster}$ from at least one another cluster ion. That cluster is considered as surviving until one of its ions leaves it (\emph{i.e.} it is not anymore located at a distance smaller than $d_\mathrm{cluster}$ from any other cluster ions) or until new ions are added to it. 

\begin{figure}[H]
\includegraphics[scale=.8]{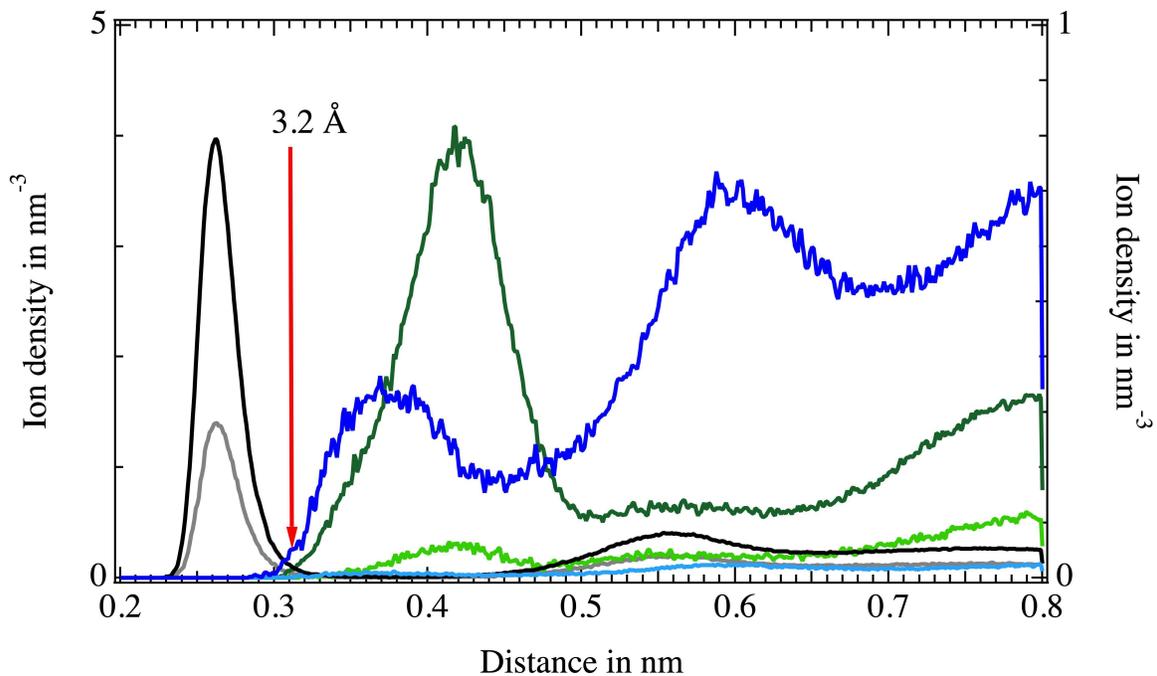}
 \caption{Ion pair radial distribution functions $g_\mathrm{ii}(r)$ within the 100k droplet bulk-like core domains. Black and grey lines: {\nacl} pairs; blue and light blue lines: \ce{[Na^+, Na^+]} pairs; green and light green lines: \ce{[Cl^-, Cl^-]} pairs. 0.2$m$ data (right axis) are shown in lighter colors than 0.6$m$ data (left axis).}
\label{fig:gdi_ion}
\end{figure}

We set the value of $d_\mathrm{cluster}$ from the inter-ionic radial distribution functions $g_\mathrm{ii}(R_\mathrm{ii})$ computed along our simulations. The three kinds of functions, corresponding to {\nacl}, \ce{[Na^+,Na^+]} and \ce{[Cl^-,Cl^-]} pairs, are shown in Figure \ref{fig:gdi_ion}. These functions agree with those computed from earlier polarizable and non-polarizable MD simulations of aqueous solutions at moderate {\nacl} concentrations, see among others Refs.~[\citenum{degreve99,uchida04,soniat16,fedkin19}]. In particular, our functions $g_\mathrm{NaCl}$ all present a first sharp peak located at 0.27~nm and which extends up to 0.32~nm, a distance at which the functions $g_\mathrm{NaNa}$ and $g_\mathrm{ClCl}$ start to be non-zero. We thus set $d_\mathrm{cluster}$ to 0.32~nm. Our cluster definition meets that proposed by Hassan~\cite{hassan08,hassan08_2} and Rick and co-workers~\cite{soniat16} to analyze {\nacl} aqueous solutions. As the first peak of $g_\mathrm{NaCl}$ corresponds to {\nacl} CIP pairs, this cluster definition allows to identify clusters in which ions are at contact rather than largely incorporating water (with ions in SSIP states, for instance). Note, however, SSIP clusters (and assemblies of clusters) are inferred to represent also a sizable proportion of the ionic assemblies that can be observed in {\nacl} aqueous solutions even at low concentrations~\cite{georgalis00,samal01,degreve99,bharmonia12,konovalov14}.

\begin{figure}[H]
\includegraphics[scale=.65]{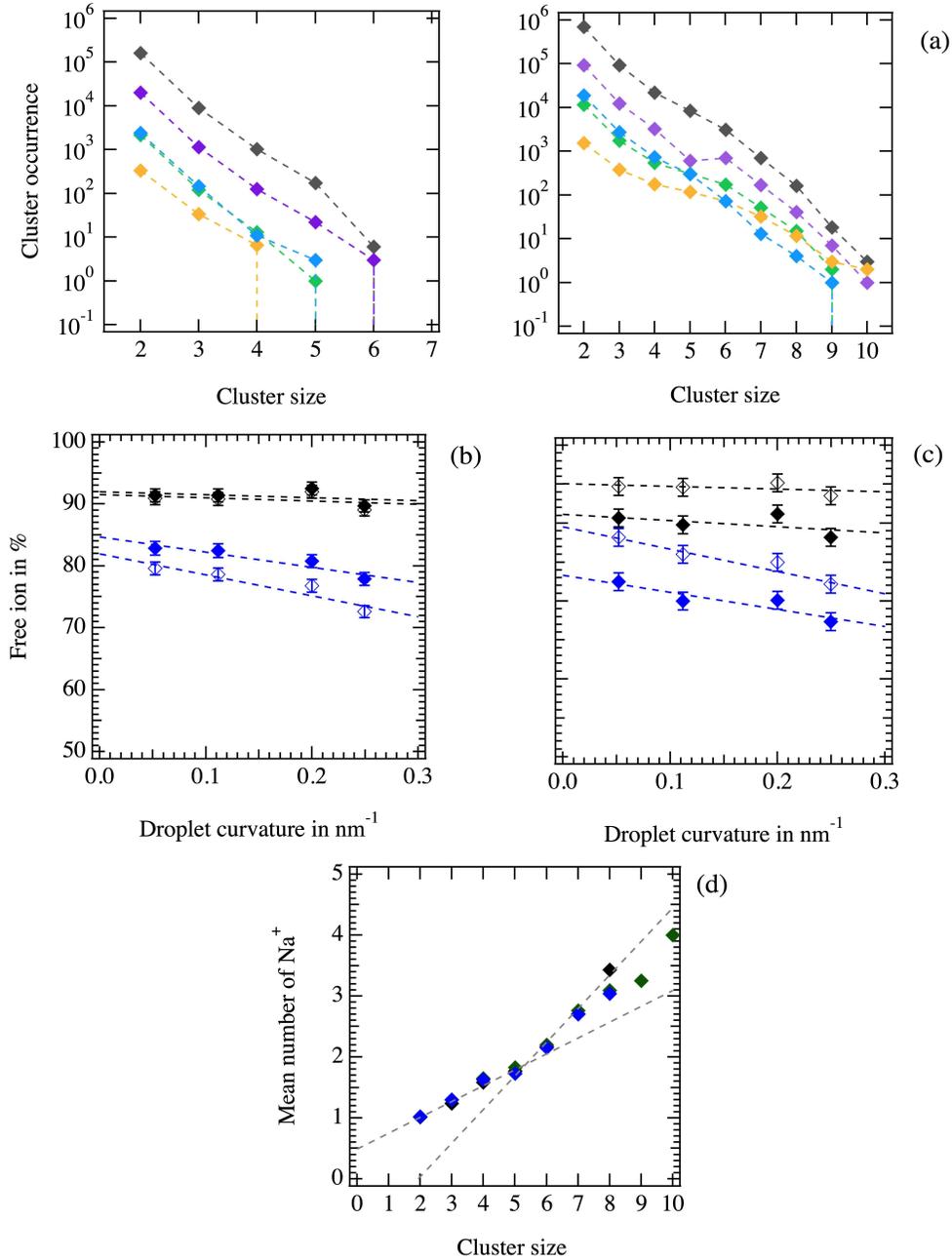}
 \caption{(a) Salt cluster distributions as a function of their size for 0.2$m$ (left) and 0.6$m$ (right) droplets. Grey, violet, blue, green and orange: data for 1M, 100k, 20k, 10k and 5k droplets. (b) Percentages of free ions within the full droplets as a function of $R_d^{-1}$. (c) Percentages of free ions within the droplet interface domains as a function of $R_d^{-1}$. For (b) and (c), full and empty symbols: {\na} and {\cl} data; black and blue symbols: 0.2$m$ and 0.6$m$ droplet data (here we consider the data of only 10k to 1M droplets). Dashed lines: linear regression fits (the regression coefficients are all > 0.95). (d) Mean number of {\na} in salt clusters as a function of their ionic size in 0.6$m$ droplets. Black, blue and green: 10k, 100k and 1M data.}
  \label{fig:clusters}
 \end{figure}

Regarding 10k-1M systems, our clustering analysis shows a noticeable presence of associated {\nacl} pairs within our droplets as well as of larger clusters comprising up to $k=$ 5 ions (0.2$m$-10k droplet) and even 10 ones (0.6$m$ droplets). However the probability of observing large clusters is weak as shown by the cluster size distributions computed from our simulations, see Figure \ref{fig:clusters}(a). These functions are computed by considering as a single event a cluster from its apparition to its 'disappearance' according to the above cluster definition rules. These functions are all fast decaying exponential functions of the cluster size and a 10-sized cluster is identified only once along the full 30~ns simulation of the 0.6$m$-1M droplet. Lastly, 3-sized clusters are relatively abundant as they represent about 10 \% of the {\nacl} pairs identified. 

In Figure \ref{fig:clusters}(b) we plot the percentage $n_\mathrm{free}^i$ of free {\na} and {\cl} ions, \emph{i.e.} ions not included in a cluster within the full 10k-1M droplets, as a function of their curvature $R_d^{-1}$. $n_\mathrm{free}^i$ increases linearly as $R_d^{-1}$ decreases for both ions, even if that quantity is almost converged for 0.2$m$ droplets as soon as 10k (within the error bars). Regarding 0.6$m$ droplets, $n_\mathrm{free}^i$(10k) data are still far from being converged compared to 1M ones. Another striking difference between 0.2$m$ and 0.6$m$ droplets is the dissymmetric behavior between {\na} and {\cl}: in 0.6$m$ droplets, the amount of free {\na} is larger than free {\cl}, by 4 (10k) to 7 (1M) \%. Such a dissymmetry also exists for 0.2$m$ droplets but the difference in free {\na} and {\cl} is less accented (about 1\%). To our opinion the dependence of the $n_\mathrm{free}^i$ quantities to the droplet radius $R_d$ may be interpreted as for the ion densities $\rho_d^i$ within the droplet bulk-like core domains, \emph{i.e.} that dependence arises from the ion exclusion shell at the close vicinity of the droplet boundary, whose extension does not depend on the droplet size. 

However the percentages of free ions within the droplet interface domains (where {\cl} ions dominate) are far to meet the percentages computed within the full droplet. In Table \ref{tab:free_abundance}, we summarized the mean quantities $n_\mathrm{free}^i$ corresponding to the droplet bulk-like core, transition layer and interface domains. These data show free {\na} to dominate over free {\cl} within the droplet core, contrary to what is observed at the vicinity of the droplet boundary where free {\cl} are more abundant than {\na} by about 5\%, regardless of the salt concentration and the droplet size.

In Figure \ref{fig:clusters}(d) we plot the mean number of {\na} per cluster as a function of the cluster size for 0.6$m$ droplets. The compositions of clusters comprising more than 2 ions are dissymmetric: they contain noticeably more {\cl} than {\na} even for even-sized clusters. That is also observed for 0.2$m$ droplets. However the relative abundance of {\na} \emph{wrt} the cluster size obeys two linear regimes: for large clusters (whose size $k$ is $\geq$ 5), the proportion of {\na} is reinforced compared to smaller clusters, even if large clusters still comprise more {\cl}. The weaker proportion of {\na} within clusters may be interpreted as arising from the slightly larger stability of the symmetric linear trimer \ce{[NaCl_2^-]} compared to \ce{[Na_2Cl^+]}, by about 3 \kcal in the gas phase, as predicted by quantum MP2/CBS computations and our force field, see Figure~S3 of \supporting. The polarizability of {\na} is one order of magnitude smaller than the {\cl} one. The electric fields arising from the two lateral ions on the central ones in both the linear trimers cancel out and that explains the relative stability of the latter trimers. 

\begin{figure}[H]
\includegraphics[scale=.7]{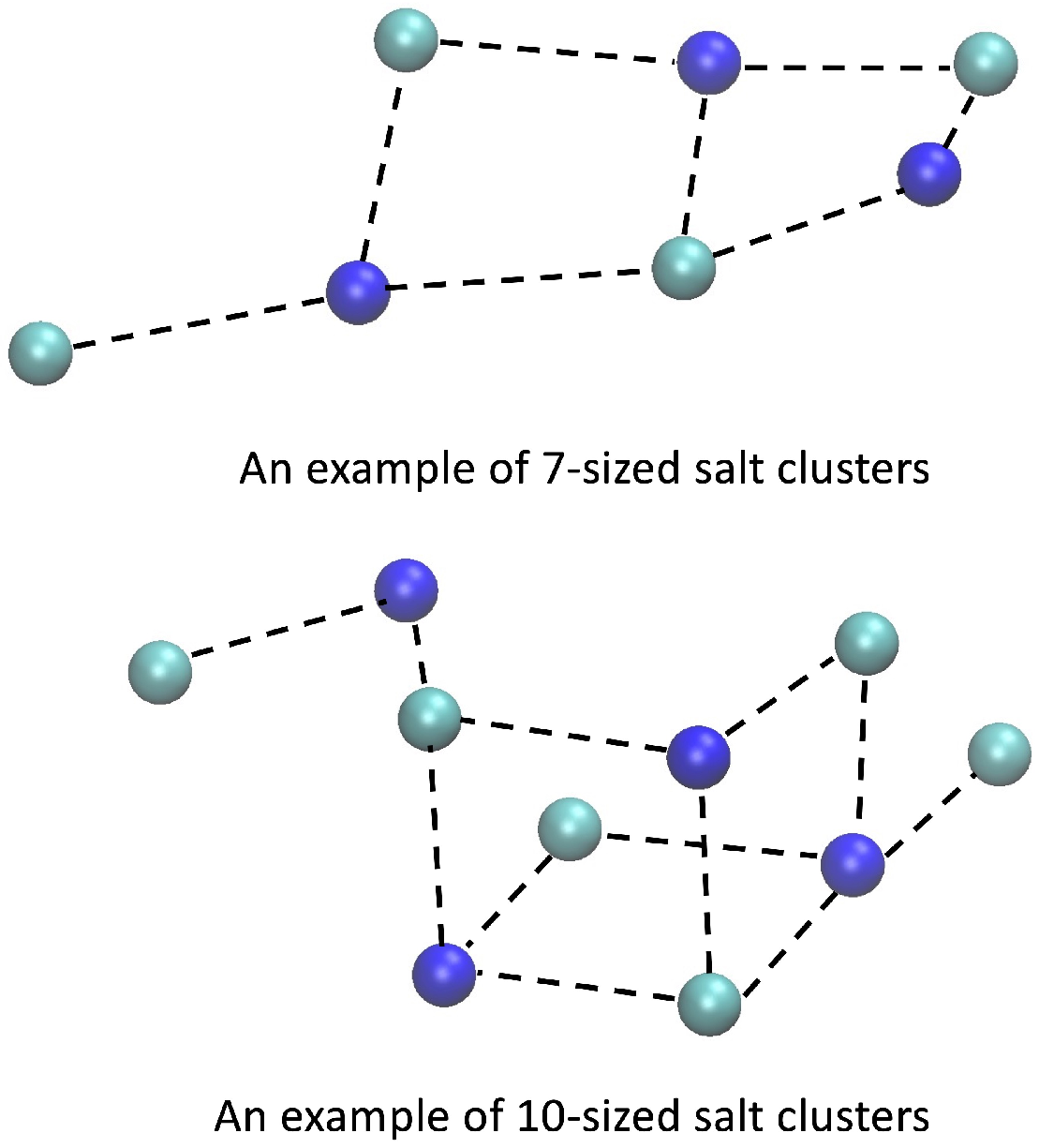}
 \caption{Examples of 7 and 10-sized salt clusters identified along our MD simulation of the 0.6$m$ 1M droplet. {\na} and {\cl} are shown in dark and light blue, respectively. Dashed lines connect ions interacting at short range, from 0.254 and 0.282~nm.}
  \label{fig:cluster_examples}
 \end{figure}

Up to relatively large sizes ($k$=7-8), the clusters are usually planar on average. For $k \geq 5$ the structure of the most abundant clusters correspond to a {\nacl} symmetric hetero tetramer to which a few more ions (preferentially {\cl}) are connected. Three-dimensional structures are observed only in 9- and 10-sized clusters: one or two {\cl} are then connected to an almost cubic and symmetric salt octamer, see the cluster snapshots shown in Figure \ref{fig:cluster_examples}, for instance. 

Our 0.6$m$ fractions of free ions extrapolated to the bulk limit (\emph{i.e.} $R_d^{-1} \rightarrow 0$) meet those reported from polarizable MD simulations of  {\nacl} 1$m$ aqueous solutions performed by Rick and co-workers~\cite{soniat16}. The force-field parameters in the latter study were refined to reproduce osmotic pressure experimental data. However contrary to our results, there is an excess of {\na} within the salt clusters identified along the 1$m$ simulations of Rick and co-workers and that yields thus a slight predominance of {\cl} among the free ions in the latter simulations. The parameters of our force field are assigned from a different \emph{ab-initio}-based strategy. However both our data and the latter one clearly support noticeable association phenomena of {\na} and {\cl} ions in aqueous solutions from low to moderate concentrations $\leq$ 1$m$, as also predicted by non-polarizable simulations, see below. However we may note the opposite conclusion drawn from MD simulations performed by means of the sophisticated reactive force field ReaxFF~\cite{fedkin19} and which predict {\nacl} to be already fully dissociated at the 1$m$ concentration. 

We estimated the mean survival time $t_s^k$ of $k$-sized clusters from the probability survival function $p_i^k (t_0,t_0 +n \tau^*)$ of a given cluster $i$ of size $k$ identified along a simulation. $\tau^*$ is the time sampling interval of our MD trajectories (set to 10 ps) and the survival function of a particular cluster is equal to one at all the  times $t_0 + l\tau^*$  ($0 \leq l \leq n)$ if that cluster is observed at all the $n$ successive sampled simulation snapshots starting at $t_0$, and zero otherwise. From the survival functions of all the $M_k$ clusters of size $k$ identified along a simulation, we compute the mean correlation function 

\begin{equation}
C_k(l\tau^*) = \Bigg \langle  \dfrac{1}{M_k} \sum_{i=1}^{M_k}  p_i^k(t_0,t_0+l\tau^*)) \Bigg \rangle_{t_0}.
\end{equation}
We show these correlation functions to obey (see Figure~S10 of \supporting):
\begin{equation}
C_k(t) = \exp (-t / t_s^k).
\end{equation}
Here, $t_s^k$ is the mean cluster survival time and its values for the most abundant clusters are reported in Table \ref{tab:cluster_mean_time}. They range from 130 to 470 $\pm$ 15 ps and the larger ones correspond mainly to the larger clusters, regardless of the salt concentration and the droplet size. The order of magnitude of our values $t_s^k$ (and their closeness) matches that computed by Hassan from non-polarizable MD simulations of {\nacl} aqueous solutions (whose concentration ranges from 0.1 to 3$m$).~\cite{hassan08} Interestingly, both the Hassan's and our $t_s^k$ estimates also match that computed for large clusters ($k \geq$ 6) in supersaturated {\nacl} aqueous solutions ($c \geq 14m$) from non-polarizable MD simulations~\cite{lanaro16}.

\begin{figure}[H]
\includegraphics[scale=.7]{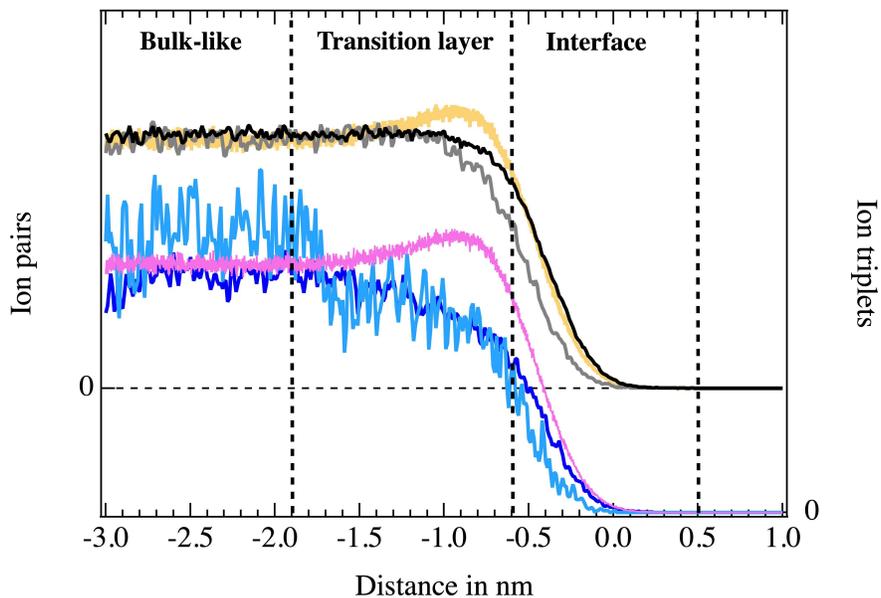}
 \caption{Normalized radial distribution functions of the salt pairs (black and grey lines, left axis) and trimers (blue and light blue lines, right axis) for 100k droplets (light colors correspond to $c=$ 0.2$m$). In light orange and violet lines: the radial distribution of {\na} for the 0.6$m$ 100k droplet for comparison purposes. Note the zero of the left and right axes are shifted for readability purposes.}
  \label{fig:cluster_distribution}
 \end{figure}

Regarding the most abundant 2 and 3-sized clusters identified along our simulations, we plot in Figure \ref{fig:cluster_distribution} their normalized radial distribution $g^{k=2,3}_\mathrm{cluster}(r)$ as a function of the distance $r$ of the cluster COM from the droplet boundary for 100k systems. The functions for 10k and 1M systems are fully in line with the latter ones. For 0.6$m$ droplets, both {\nacl} pairs and 3-sized clusters are repelled from the droplet surface as individual {\na}, and that behavior is more accented for 0.2$m$ droplets. Salt clusters are thus mainly observed within droplet bulk-like core domains. We also note a large diffusion of salt clusters across the droplet core, transition layer and interface domains: up to 90\% of the salt clusters originally formed within the transition layer and interface domains vanish or incorporate new ions in a different droplet domain.

Regarding the smallest 5k droplets, most of the results discussed above for larger systems are still valid. However we note much weaker percentages of free ions in 5k droplets, percentages that drop  to 80 \% (0.2$m$) and even 55-60 \% (0.6$m$), regardless of the droplet domain and ion type, see Table \ref{tab:free_abundance}. We note even a stronger probability to observe large clusters (up to 10 sized ones) within the 5k 0.6$m$ droplet than within larger systems, see Figure \ref{fig:clusters}(a). To our opinion that arises from the proportionally weaker ion 'accessible' volume in 5k droplets compared to larger systems (see above), which favors ion clustering. Accounting also for the ion surface excess data regarding the 0.6$m$ 5k droplet, the ion behavior within droplets appears to obey two regimes, the first, that we term as  'curvature linear' regime, for droplets whose molecular size is as small as 10k and a more complex regime for smaller droplets.

\subsection {Surface potential}

As our systems are all quasi spherical, we solve the Poisson's equation to estimate the electrostatic potential $\Phi$ at a distance $r$ from the droplet COM according to

\begin{equation} \label{eqn:phi_d}
\Phi (r) = -\frac{1}{\epsilon_0}\int_{0}^{r} \frac{C_q(r) + C_p(r)}{r^2} dr.
\end{equation}

\begin{figure}[H]
\includegraphics[scale=1.0]{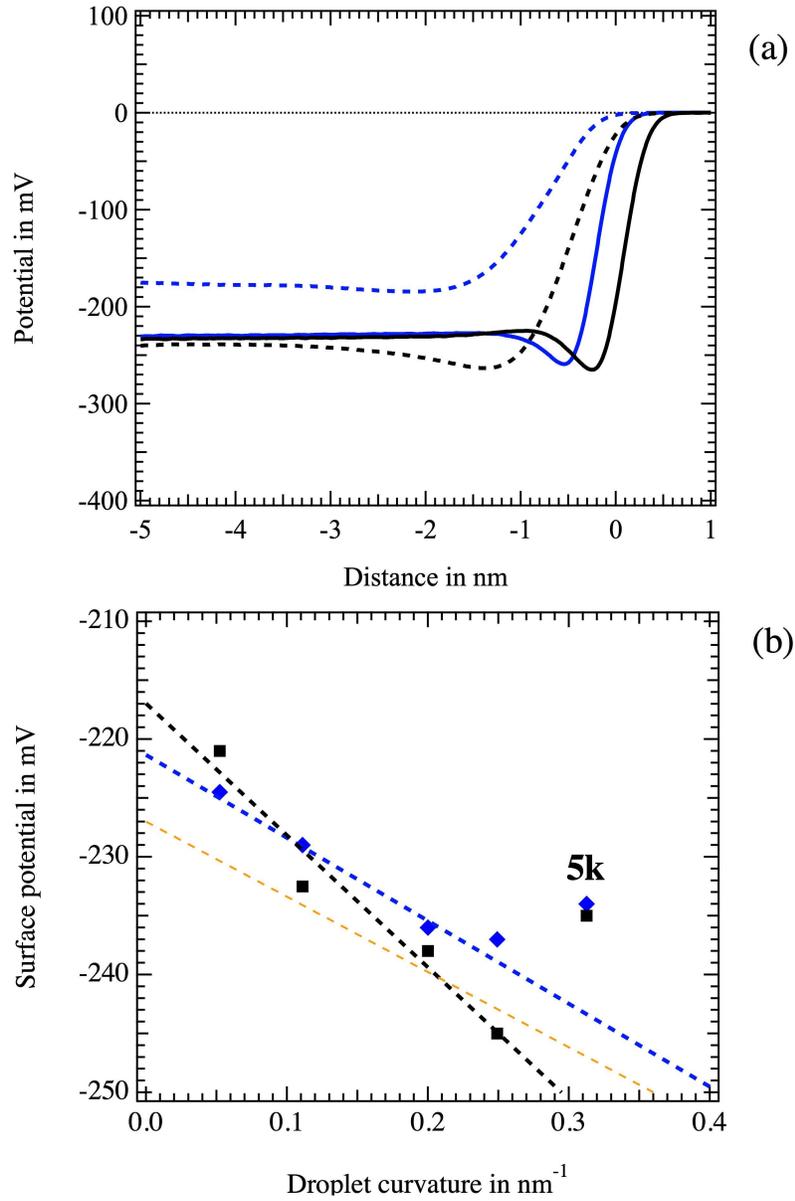}
 \caption{(a) Surface potentials $\Phi(r)$ as a function of the distance $r$ from the droplet boundary for 100k systems. Black and blue: 0.6 and 0.2$m$ data, respectively. Full and dashed lines: full surface potentials and their induced dipole components. (b) Surface potential $\Delta \Phi$ as a function of the droplet curvature $R_d^{-1}$ (in~nm$^{-1}$). Black and blue: 0.6 and 0.2$m$ data. Dashed lines: linear regression fits from data corresponding to droplets larger than 5k (in orange our former result for neat water~\cite{houriez19}).}
   \label{fig:phi}
 \end{figure}

$C_q(r)$ is the mean sum of the static charges included in a sphere of radius $r$ and $C_p(r)$ is the mean density at distance $r$ of the induced dipole projections in the radial direction. We discretely computed these functions from our simulation data with $dr$ set to 0.01~nm. The $\Phi(r)$ plots for 100k droplets are shown in Figure \ref{fig:phi}(a). The surface potential $\Delta \Phi$ of our systems is the difference in the $\Phi(r)$ values corresponding to $r = R_d -2$ and $ R_d + 1$~nm.

For 10k-1M salty droplets and as already reported for pure water systems~\cite{houriez19}, the surface potentials $\Delta \Phi$ of all our systems are linearly correlated to the droplet curvature $R_d^{-1}$, see  Figure \ref{fig:phi}(b). The extrapolated  planar interface values $\Delta \Phi(\infty)$ are almost equal : -221 (0.2$m$) and -217 (0.6$m$) $\pm$ 3~mV. They are close to our $\Delta \Phi$ estimate for neat water as using our water model TCPE/2013: -227~mV~\cite{houriez19}. Note the $\delta_\Phi$ length is larger for 0.6$m$ droplets than for the 0.2$m$ ones, 0.32 and 0.51 $\pm$ 0.07~nm, respectively (for neat water $\delta_\Phi = 0.14$~nm~\cite{houriez19}).

According to Equation (\ref{eqn:phi_d}) the surface potentials $\Delta \Phi$ are the sum of the static charge and induced dipole components $\Delta \Phi_q$ and $\Delta \Phi_p$. For pure water systems, the latter components are negative and their ratio is about 3:1~\cite{houriez19}. For our \ce{NaCl} droplets,  the $\Delta \Phi$ values are dominated by the induced dipole components $\Delta \Phi_p$, regardless of the salt concentration.  

Regarding neat water and 1.1-2.1$m$ \ce{NaCl} aqueous solutions, Ishyama and Morita~\cite{ishiyama07} also decomposed the surface potentials in their $\Delta \Phi_q$ and $\Delta \Phi_p$ components from MD data generated using an induced dipole-based polarizable force field whose parameters were assigned to reproduce experimental data regarding neat water and electrolyte aqueous solutions at ambient conditions. Contrary to our results, the latter authors showed the ratio of the $\Delta \Phi_q$ and $\Delta \Phi_p$ components of their salty solutions to be about 3:1, \emph{i.e.} the $\Delta \Phi_q$ contributions  are thus dominating. However in line with our 0.6$m$ data, the latter authors concluded the surface potential $\Delta \Phi$ of neat water and of their salty solutions to be equal (within 6~mV), even if their $\Delta \Phi$ values are about twice as large as our own. 

Taking in mind the overall large variability of the theoretical values $\Delta \Phi$ reported for neat water from both polarizable and non-polarizable force fields (see among others the values reported in Ref.~\citenum{lamoureux06}), to our opinion the differences between our $\Delta \Phi$ values and those reported by Ishyama and Morita arise from the different strategies used to adjust the force-field parameters. However both our force field and that of the latter authors predict the same trends for the change in the $\Delta \Phi$ values from neat water to low/moderately concentrated salty \ce{NaCl} aqueous solutions, \emph{i.e.} the presence of salt has a weak effect on the surface potential of aqueous solutions. Experimentally, Jarvis and Scheiman~\cite{jarvis68} reported differences in $\Delta \Phi$ values between \ce{NaCl} aqueous solutions (whose concentrations $c$ are < 1$m$) and neat water to be about +1~mV whereas a recent experimental study of Allen and co-workers showed that difference to be +120~mV for $c=1m$~\cite{adel21}. Both our estimates of $\Delta \Phi$ and the earlier ones of Ishyama and Morita support thus the earlier experimental estimate of Jarvis and Scheiman. 

Regarding 5k droplets and regardless of $c$, their  $\Delta \Phi$ values differ noticeably from the expected values extrapolated from the data of larger systems, see Figure \ref{fig:phi}(b). Accounting also for the structural and ion clustering data discussed above, the properties of salty 0.2 and 0.6$m$ droplets thus obey a linear regime of the droplet curvature down to a molecular size included between 5k and 10k. That corresponds to a droplet radius of about 3 nm.

\subsection{Water vapor density and the Kelvin term}

The Kelvin term is  one of the main components of the K\"ohler's equation \cite{koehler36} commonly used to discuss the stability of SSAs in the atmosphere and the formation of cloud condensation nuclei (see among others Ref.\citenum{farmer15}). The historical (standard) Kelvin term relates the equilibrium vapor pressure $p_d$ to the radius $R_d$ of a spherical liquid droplet according to
\begin{equation} \label{eqn:kelvin}
p_d = p_{\infty} \exp \left(   \dfrac{2 \gamma _\infty }{ \mathrm{RT} \rho _\mathrm{\infty} R_d}  \right).
\end{equation}
Here, $p_\infty$ and $\gamma_\infty$ are the saturation pressure and the surface tension of the liquid (planar interface)/gas phase equilibrium, respectively. $\rho_\infty$ is the density of the liquid phase, and R is the ideal gas constant. Assuming the water vapor at ambient conditions to be an ideal gas, we may rewrite the Kelvin term in a logarithmic form as
\begin{equation}
\ln \rho_d^g  =   \ln{\rho_\infty^g} + \dfrac{2 \gamma _\infty }{ \mathrm{RT} \rho _\infty} \times \dfrac{1}{R_d},
\end{equation}
here, $\rho_d^g$ and $\rho_\infty^g$ are the water vapor densities corresponding to the droplet/vapor and liquid water/vapor equilibrium, respectively. Simulations based on our force field predict $\gamma_\infty$ to amount to 63.3, 64.3 and 64.4 $\pm$ 0.1 mN m$^{-1}$ for neat water, 0.2$m$ and 0.6$m$ \chem{NaCl} solutions at ambient conditions, respectively (see \supporting). These values are about 10 \% smaller than experiment, but they lie within the range of values reported from earlier simulations based on different force fields \cite{neyt13}. However, and at the difference of many of these earlier simulations, our data predict weak positive increments of the surface tension from neat water to salty solutions (at weak and moderate concentrations) in agreement with experiment  \cite{johansson74}.

Because of the sub-micron scale of our droplets, it is more reasonable to consider a modified version of the standard Kelvin term based on the droplet surface tension $\gamma_d$ that obeys the Tolman relation \cite{alekseechkin18}

\begin{equation} \label{eqn:tolman}
\gamma_d = \gamma_\infty \left(  \dfrac{R_d}{R_d + 2 \delta_\gamma} \right).
\end{equation}

Here $\delta_\gamma$ is the Tolman length. Neglecting the weak curvature dependence of the water density within the droplets core (see above), that yields the water vapor densities to obey

\begin{equation} \label{eqn:water_vapor_density}
\ln \rho_d^g  =   \ln{\rho_\infty^g} + \dfrac{2 \gamma _\infty }{ \mathrm{RT} \rho _\infty} \times \dfrac{1} {(R_d + 2 \delta_\gamma)}.
\end{equation}

As the cavities in which our droplets lie are large enough, we computed from our simulations the water mean molecular densities $\rho_d^g$  of  the vapor phase surrounding the droplets within a spherical shell extending from +1 to + 2~nm from the droplet surfaces. Water radial distribution functions $\rho(r)$ within those shells are plotted in Figure~S14 of \supporting as a function of the distance $r$ from the surface for 10k to 1M salty droplets. Those plots show converged $\rho_d^g$  densities that decrease from about 1.0 down to about 0.5 10$^{-3}$ molecule~nm$^{-3}$ as the droplet size increases. Those density values agree with the vapor density corresponding to liquid water, about 0.8 10$^{-3}$ molecule~nm$^{-3}$ from experimental data at ambient conditions. We computed the uncertainty of our mean vapor density from a standard block analysis method (see Figures~S12 and~S13 of \supporting). For droplets simulated at the 100~ns scale and above (5k to 100k), that uncertainty is about 8\%, whereas it amounts to about 16\% for the 1M droplets that were simulated at the 30~ns scale. Note also that we performed multiple simulations of 1k pure aqueous droplets at the 500~ns scale: the water vapor densities computed along all these simulations agree within the error bars computed from the standard block analysis method.

\begin{figure}[H]
\includegraphics[scale=.8]{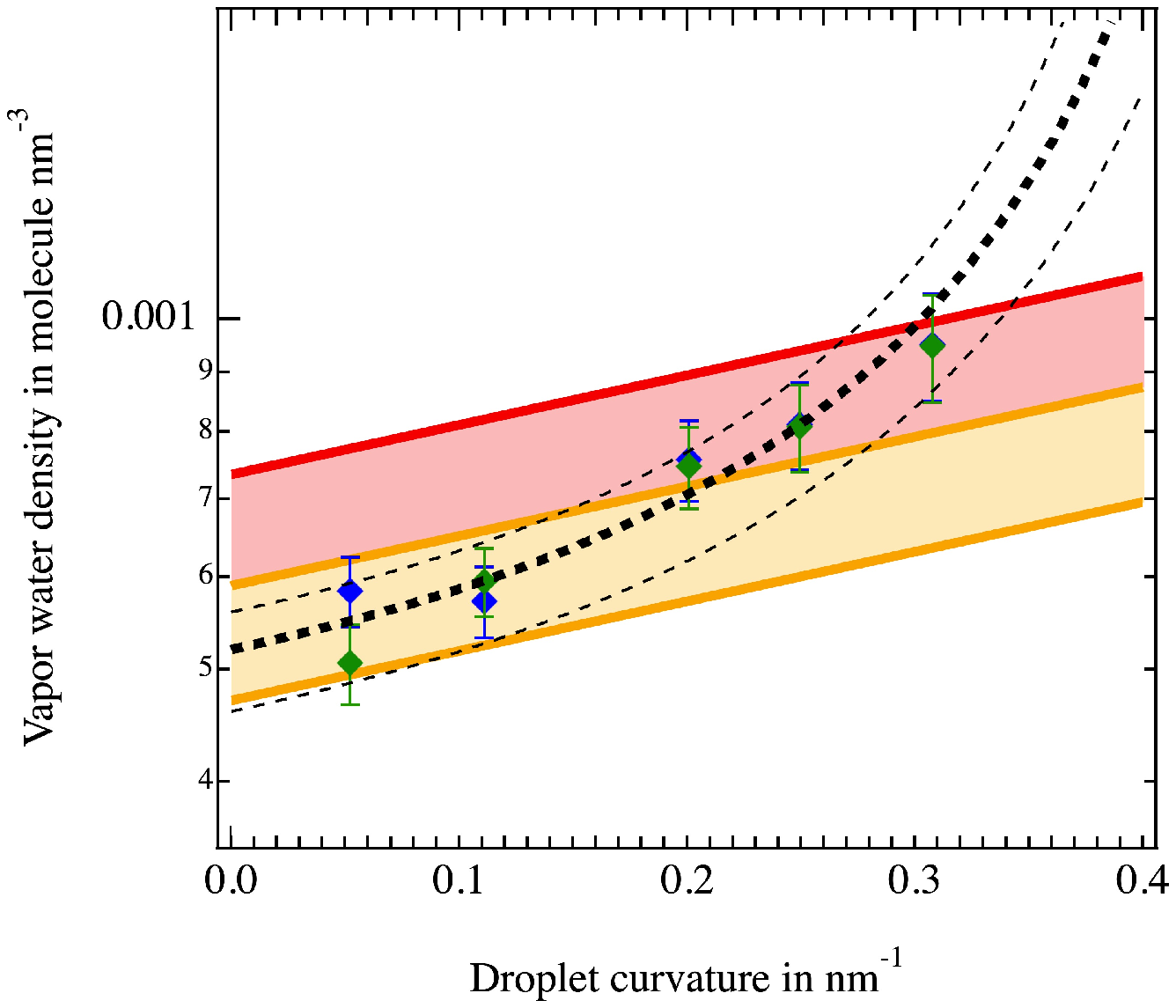}
 \caption{Water vapor density (ln scale) as a function of the droplet curvature $R_d^{-1}$. Green and blue diamonds: 0.2 and 0.6$m$ data. Bold dashed line: best interpolation of the raw data according to relation (\ref{eqn:kelvin}) and using a Tolman length  $\delta_\gamma$ set to -0.85~nm. Thin dashed lines: best interpolations of the data accounting for the uncertainties (upper line, $\delta_\gamma$ = -0.9~nm, lower line: $\delta_\gamma$ = -0.8~nm). Red and orange bold lines: standard Kelvin terms ($\delta_\gamma$ = 0) to reproduce 5k (red), 20k (upper orange) and 100k (lower orange) data.}
   \label{fig:tolman}
 \end{figure}

The plots of the quantities $\ln (\rho_d^g)$ as a function of the droplet curvature $R_d^{-1}$ reported in Figure \ref{fig:tolman} show the water vapor densities $\rho_d^g$ to obey (within the error bars) a Kelvin term based on droplet surface tensions $\gamma_d$ corresponding to a negative Tolman length at the 1~nm scale, \emph{i.e.} $\delta_\gamma=$ -0.85 $\pm$ 0.05~nm, regardless of the salt concentration. We also investigated  pure water droplets whose molecular size ranges from 1k to 20k according to our modeling protocol. The simulations were performed at the 500~ns (1k-2k), 200~ns (5k-10k) and 100~ns (20k) scale. In agreement with all the theoretical data reported to date \cite{joswiak13, factorovich14,leong18}, our simulations suggest for pure water droplets a weak and a priori negative Tolman length $\delta_\gamma$ (at least -0.1~nm, see Figure~S11 of \supporting). 

The vapor densities $\rho_\infty^g$ corresponding to neat water and salty aqueous planar interfaces that allow us to best reproduce our $\rho_d^g$ data amount to about 0.52 (salty solutions) and 0.64 (neat water) 10$^{-3}$ molecule~nm$^{-3}$. Our simulations predict thus large \chem{NaCl} SSAs to be stable in the atmosphere at a relative humidity of about 80 \%, in agreement with experiment \cite{bertram18}. Along our simulations of planar interfaces, we also computed the corresponding water vapor densities $\rho_\infty^g$. Their order of magnitude agrees with the above extrapolated estimates. However these data suffer from large uncertainties (about 30-40\% of the $\rho_\infty^g$ values, see Figure~S14 of \supporting). To our opinion that arises from the weak extension of the surfaces (and from the overall large volume of the vapor phase) of the simulated planar interfaces. Compared to them, the boundary surfaces of 5k to 1M  droplets (and thus the number of water molecules able to escape from the droplet surface) are from 16 up to 500 times larger. 

\section{Conclusion}

We have investigated at ambient temperature {\nacl} salty aqueous droplets whose water molecular sizes vary from 1k to 1M and whose salt concentrations are 0.2$m$ (brackish water) and 0.6$m$  (sea water) using a polarizable \emph{ab initio}-based force field and MD simulations at the 500 (1k/2k), 200 (5k/10k), 100 (100k) and 30 (1M)~ns scale. Because of their size and composition, 100k and 1M salty droplets  correspond to ideal sub-micron SSAs as being produced by oceans.

Regarding large systems (from 5k to 1M), our simulations show {\nacl} droplets to be organized in three main domains: the interface domain (at the vicinity of the droplet boundary and where {\cl} dominate within the droplet interior sub-region), then the transition layer (where {\na} dominate at its upper boundary) and finally the bulk-like core domain corresponding to a homogeneous salty solution. Interestingly, the spatial extension of these domains does not depend on the droplet size and composition. By measuring the distances from the droplet boundary, the interface domain extends from +0.5 to -0.6~nm (with a maximum of the {\cl} over concentration relative to {\na} at -0.3~nm), the transition layer extends from -0.6 down to -1.9~nm (with a maximum of the {\na} over concentration relative to {\cl} at -0.9~nm), and the bulk-like core domain starts at -1.9~nm.  Compared to earlier simulations the most striking discrepancies concern the composition of large {\nacl} aggregates that may be observed in aqueous environments: our droplet simulations show {\cl} to dominate in all the most probable aggregates larger than dimers whereas earlier bulk simulations also based on a polarizable force field but adjusted from experimental data predict the contrary. As the parameters of our force field are assigned only from quantum data that show the isolated linear trimer \ce{[(Cl^{-})_2,Na^+]} to be more stable than \ce{[(Na^{+})_2,Cl^-]}, that clearly suggests the need of accounting for quantum data regarding small molecular aggregates while building a force field to reproduce macroscopic experimental properties.

Contrary to ion spatial distributions, the magnitude of the properties $A$  that we investigated (like surface potential, ion surface excess and ion droplet-core densities) for large droplets are linearly correlated to the droplet curvature $R^{-1}$ according to 

\begin{equation} 
A = A_\infty \left(1- \dfrac{2 \delta_A} {R}\right) + O(\dfrac{1}{R^2}).
\end{equation}
Here $A_\infty$ is the extrapolated bulk (planar interface) value of the quantity $A$ and $\delta_A$ is a length whose sign and magnitude measure the surface curvature effect on $A$. Regarding the reliability of our simulations, we note all the extrapolated data $A_\infty$ estimates to be in line with or included within most of the available experimental and theoretical data.
Regarding the magnitudes of the lengths $\delta_A$ they amount from a few tenths of~nm up to at most a few~nm. Interestingly, our simulations show the surface tension of NaCl salty droplets to obey the Tolman relation with a Tolman length negative and at the 1~nm scale, whereas our simulations of 1k to 20k pure aqueous droplets yield a much weaker Tolman length (about -0.1~nm) for neat water droplets in agreement with earlier studies based on different molecular modeling approaches  \cite{joswiak13,factorovich14,leong18}. Lastly the limit of validity of the above relation corresponds to droplets whose size is included between 5k and 10k, \emph{i.e.} droplets whose radius is about 3-4 nm.

Another important difference between pure aqueous and salty droplets is the stability of small droplets whose molecular size is lower than 5k. Our simulations show salty systems to present a source of instability arising from the lack of a homogenous ionic domain at their center contrary to larger systems. To our opinion that explains the splitting of 1k and 2k  0.6$m$ droplets in smaller subsystems at the early stages of multiple independent simulations, whereas 1k/2k pure water droplets are stable along multiple 500~ns scale  simulations (\emph{i.e.} they correspond to quasi-spherical liquid nanodrops whose radius is constant along the simulations, on average).

The uncertainties regarding our data prevent us to further investigate the dependence of droplet properties to higher order curvature terms $R^{-n}$ ($n >1$). In all our data regarding 10k to 1M salty droplets show their properties to be usually well reproduced using a standard and simple linear relation of the droplet curvature. Regarding the surface tensions of our salty droplets, they a priori obey the Tolman relation. However as the magnitude of the Tolman length $\delta_\gamma$ is about 1~nm and because of the uncertainties regarding water vapor densities, our surface tension data may also obey to more complex laws as those discussed in Refs.[\citenum{troster12,bruot16}].

Our 10k to 1M salty droplets corresponds to ideal sub-micron SSAs that might dominate the aerosol distributions at marine boundary layers \cite{bates98,zhang14}. Regarding atmospheric science, our surface tension results support the validity of the Kelvin relation in its droplet form, see Eq. (\ref{eqn:kelvin}), to model salty SSAs whose radius is as small as 3-4~nm. Together with the water activity $a_w$, the Kelvin term is the second component of the K\"ohler equation relating the ambient relative humidity RH to the SSA radius $R_d$ according to

\begin{equation}
\mathrm{RH} = a_w \times \exp \left(  \dfrac{2\gamma_d}{\mathrm{RT} \rho_l R_d} \right).
\end{equation}
 
Assuming a standard relation for the water activity $a_w$ \cite{farmer15}, the effect of a negative Tolman length at the 1~nm scale yields to weaken the number of sub-micron NaCl SSAs that might activate for RH values included between 1.003 and 1.01 (and thus evolve towards stratocumulus and warm cumulus clouds as usually inferred \cite{farmer15}) by a few percents as compared to the prediction of a standard K\"ohler equation based on infinite planar surface tension data (see Figure~S15 of \supporting). However marine sub-micron scale SSAs are enriched by organic materials that may alter their properties as compared to the ideal salty droplets that we studied here. We are presently simulating salty aqueous SSAs in presence of fatty acids by means of new polarizable force fields  \cite{ozgurel22} to further assess the behavior of more complex sub-micron SSAs.

% Supplementary Material

\section*{Supporting Information}

All the plots mentioned as supplementary materials in the manuscript are provided as \supporting. Details regarding the computations of surface tensions from planar interface simulations, the short range many body functions to improve the modeling of ions in aqueous environments as well as the computations of cumulative condensation nuclei from the K\"olher relation are also provided as \supporting. Lastly, the raw data computed along our simulations of the ion and salt clusters (from dimers to pentamers) radial distribution functions, as well as of the functions $C_q(r)$ and $C_p(r)$ to compute the surface potentials are also provided as \supporting.

% Acknowledgement

\section*{Acknowledgments}

This work was granted access to the TGCC HPC resources under the Grand Challenge allocation [GC0429] made by GENCI. We acknowledge support by the French government through the Program "Investissement d'avenir" (LABEX CaPPA / ANR-11-LABX-0005-01 and I-SITE ULNE / ANR-16-IDEX-0004 ULNE), as well as by the Ministry of Higher Education and Research, Hauts de France council and European Regional Development Fund (ERDF) through the Contrat de Projets État-Région (CPER CLIMIBIO). We further acknowledge the "Groupement de recherche" GDR 2035 SolvATE. 

% Tables
\newpage

\begin{table}[H]
\footnotesize
\begin{center}
\begin{tabular}{cccccccccccc}
\toprule
 Droplets & $c$ & $N_\mathrm{pair}$ & $R_d$  & $\eta_d$ & $\rho_d$ & $\Gamma_\chem{Na^+}$ & $\Gamma_\chem{Cl^-}$ & $\rho_d^\chem{Na^+}$ &   $\rho_d^\chem{Cl^-}$ \\
\midrule
 5k & 0.2$m$ &  15         &   3.2     &  0.195   &  33.75   & -0.0691 & -0.0629 & 0.0161 & 0.0167  \\
        & 0.6$m$ & 49        &   3.2     &  0.192   & 33.29    & -0.2218 & -0.2180 & 0.0542 & 0.0539  \\
 10k & 0.2$m$ &  37         &   4.0     &  0.205   & 33.63     & -0.0691 & -0.0668 & 0.0185 & 0.0184  \\
        & 0.6$m$ & 102        &   4.0     &  0.205   & 33.29     & -0.1204 & -0.1276 & 0.0486 & 0.0491  \\
20k & 0.2$m$ &  53         &   5.0     &  0.215   &  33.63    & -0.0526 & -0.0502 & 0.0129& 0.0122  \\
        & 0.6$m$ & 176        &   5.0     &  0.212   &   33.27   & -0.1361 & -0.1313 & 0.0436 & 0.0448  \\
 100k & 0.2$m$ & 398      &   8.9     &  0.243   & 33.39     & -0.0572 & -0.0576 & 0.0152 & 0.0152   \\
         & 0.6$m$ & 1038     &  9.0      &  0.239   & 33.17     & -0.1214 & -0.1245 & 0.0380 & 0.0381   \\
 1M & 0.2$m$ &  4000      & 19.2     &  0.254   & 33.33     & -0.0561 & -0.0566 & 0.0141 & 0.0143   \\
         & 0.6$m$ & 10000   & 19.2     &  0.253   & 33.12     & -0.1136 & -0.1137  & 0.0356 & 0.0357   \\
\bottomrule
\end{tabular}
\end{center}
\caption{Main droplet properties. $N_\mathrm{pair}$: number of ion pairs in 0.2 and 0.6$m$ systems. $R_d$ and $\eta_d$: mean droplet radius and thickness (in~nm). $\Gamma_{\chem{Na^+/Cl^-}}$: ion surface excess in~nm$^{-2}$. $\rho_d$ and $\rho_d^{(\chem{Na^+,Cl^-})}$: mean water and ion densities within the droplet bulk-like core domain (expressed in molecules(ions)~nm$^{-3}$).}
\label{tab:droplet_data}
\normalsize
\end{table}

\newpage

\begin{table}[H]
\footnotesize
\begin{center}
\begin{tabular}{ccccccc}
\toprule
 Size &  \multicolumn{2}{c}{bulk-like core} & \multicolumn{2}{c}{transition layer} & \multicolumn{2}{c}{interface}\\
\cmidrule(lr){2-3}\cmidrule(lr){4-5}\cmidrule(lr){6-7}
           & {\na} & {\cl}  & {\na} & {\cl}  & {\na} & {\cl} \\
\midrule
 0.2$m$ \\
 5k & 80.6 & 77.8 & 83.0 & 76.8 & 74.2 & 86.5 \\
 10k & 90.0 & 89.0 & 90.6 & 87.6 & 88.2 & 93.0 \\
 20k & 92.8 & 91.7 & 92.8 & 90.9 & 91.0 & 95.4 \\
 100k & 91.3 & 90.6 & 91.8 & 90.0 & 89.8 & 94.6    \\
 1M & 91.3 & 90.8 &  92.0 & 90.6 & 90.7 & 94.7 \\
 \midrule
 0.6$m$ \\
 5k    & 60.1 & 46.0 & 65.2 & 47.6 & 63.9 & 53.0 \\
  10k & 76.9 & 69.6 & 79.8 & 70.0 & 78.3 & 82.9 \\
 20k &79.9 & 75.1& 81.4 & 72.9 & 79.9 & 84.4 \\
 100k &   81.8 & 77.5 & 83.8 & 77.7 & 82.9 & 86.6 \\
 1M & 82.6 & 79.2 & 84.1 & 79.4 & 82.5 & 88.2 \\
\bottomrule
\end{tabular}
\end{center}
\caption{Percentages of free ions in the three main droplet domains, in \%. Their mean uncertainty is $\pm$ 2 \%.}
\label{tab:free_abundance}
\normalsize
\end{table}

\newpage

\begin{table}[H]
\footnotesize
\begin{center}
\begin{tabular}{cccc}
\toprule
      Cluster size      & 2 & 3 & 4 \\
\midrule
 0.2$m$ \\
 5k & 148 & 158 & 120 \\
 10k &  140 & 215& 104 \\
 20k & 135 & 215 & 470 \\
 100k & 130  & 197  & 330    \\
 1M &  137 & 190 & 249 \\
\midrule
 0.6$m$ \\
 5k & 144 & 172 & 213 \\
  10k &  133 & 175 & 291\\
 20k & 126 & 176 & 352 \\
 100k &  130 & 187 & 326 \\
 1M &  136 & 202 & 359 \\
\bottomrule
\end{tabular}
\end{center}
\caption{Mean survival times $t_s^k$ of the most abundant salt clusters of size $k=2,3$ and 4. All values in ps. Their mean uncertainty is $\pm$ 15 ps.}
\label{tab:cluster_mean_time}
\normalsize
\end{table}

% Figures

\newpage

\newpage

\newpage

\newpage

\newpage

\newpage

\newpage

\newpage

\newpage

\newpage

\newpage

% Bibliography

\clearpage
\bibliography{bibliography}

% End of the document
\end{document}

\newpage

%\linenumbers

% Test Area Methods

\section*{Short range many body functions to improve the force field reliability regarding large ion/water clusters : their effect on NaCL PMF in neat water}

To improve the accuracy of our force fields to model
\begin{itemize}
\item the large hydrated halide anion clusters (comprising more than 6 water molecules) that can be observed in aqueous environments (up to 12 water molecules can accommodate in the first hydration shell of halides and carboxylate anions);
\item alkali/water clusters presenting a partial or complete second hydration shell,
\end{itemize}
we built up specific short range many-body functions (denoted as $U^{sc}$), whose analytical form is close to the $U^{coop}$ one to model hydrogen bonds. These functions $U^{sc}$ improve the modeling of water/water interactions only at the close vicinity of anionic and cationic centers. For the present purpose, we used the $U^{sc}$ functions (and their parameters) allowing one to model the hydration of \ce{Na^+} and \ce{Cl^-} from our earlier studies~\cite{real19,houriez19}. The $U^{sc}$(\ce{Cl^-}) and $U^{sc}$(\ce{Na^+}) functions destabilize and stabilize the ion/water interactions in large hydrated ionic clusters, respectively.

\begin{figure}[H]
\includegraphics[scale=.7]{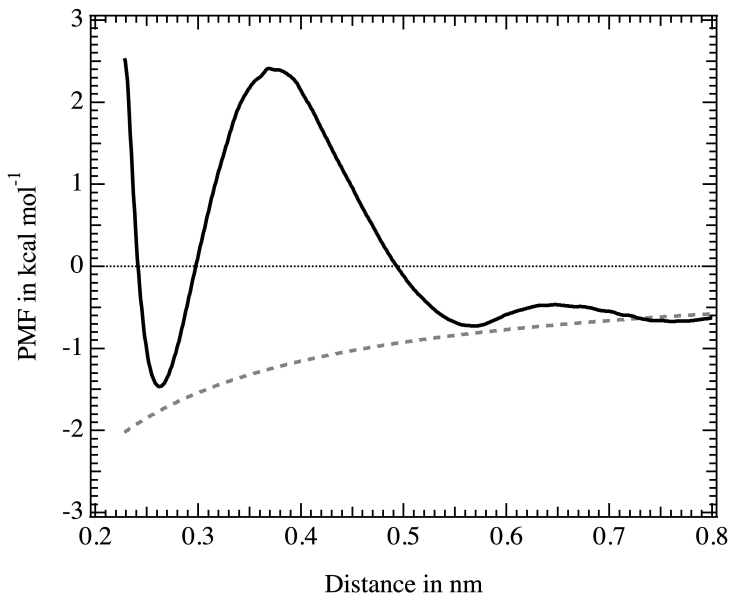}
 \caption{{\nacl} ion pair PMFs in neat liquid water at ambient conditions as computed from our force fields. Black (PMF$^\mathrm{full}$) and green lines (PMF$^\mathrm{nosc}$): PMFs computed by accounting for the both the correction potentials  $U^{sc}$(\ce{Cl^-}) and $U^{sc}$(\ce{Na^+}) and for none of them, respectively. Blue (PMF$^\mathrm{nosc(\ce{Na^+})}$) and red lines (PMF$^\mathrm{nosc(\ce{Cl^-})}$): PMF computed by accounting only for $U^{sc}$(\ce{Na^+}) and for  $U^{sc}$(\ce{Cl^-}), respectively. In dashed grey lines, the expected Coulombic PMF corresponding to two point charges of opposite sign in bulk water at ambient conditions. All the simulated PMFs are shifted to meet the Coulombic PMF value at 0.8~nm.}  \label{fig:SI-pmf}
 \end{figure}

We plot in Figure \ref{fig:SI-pmf} the {\nacl} PMF in liquid water computed from our polarizable force-field approach, as well as the PMFs computed by omitting both the correction potentials $U^{sc}$(\ce{Cl^-}) and $U^{sc}$(\ce{Na^+}) and only a single one. The latter four PMFs are denoted PMF$^\mathrm{full}$, PMF$^\mathrm{nosc}$, PMF$^\mathrm{nosc(\ce{Na^+})}$ and PMF$^\mathrm{nosc(\ce{Cl^-})}$, respectively. All our PMFs agree with the standard locations of the CIP, TS and SSIP states. Not accounting for both the correction potentials  $U^{sc}$(\ce{Cl^-}) and $U^{sc}$(\ce{Na^+}) yields a PMF that differs noticeably from those computed by accounting for these potentials as well as from the CPMD and Drude-based ones discussed in the main manuscript. In that case the energy depth corresponding to CIP is -2 \kcal, for instance. Neglecting only the correction potential $U^{sc}$(\ce{Na^+}) yields to largely over stabilize the first minium CIP (its energy depth is as low as -3.2 \kcal) whereas neglecting only $U^{sc}$(\ce{Cl^-}) yields a PMF profile that agrees the Drude-based one. These results are in line with the expected effects of our correction potentials: $U^{sc}$(\ce{Cl^-}) is built to destabilize the interactions among the water molecules in the chloride first hydration shell (that yields to favor ion pairing), whereas $U^{sc}$(\ce{Na^+}) aims at reinforcing the interaction among water molecules of the {\na} first and second hydration shell (and that disfavors ion pairing).

 \newpage

\section*{Computing the surface tension of planar interfaces}

To simulate planar liquid/gas phase interface regarding neat water and  0.2 and 0.6$m$ NaCl aqueous solutions, we first generated a 10~ns  NPT MD runs of a box comprising about 2000 water molecules and whose dimensions are $(L_x=L,L_y=L,L_z=2L)$, using periodic boundary conditions at ambient conditions according to the protocol detailed in the main manuscript. The average dimensions $\bar{L}$ corresponding to the last 5~ns segment of the trajectories are computed and new 10~ns NVT simulations are performed from the final snapshots of the NPT trajectories and the new cell dimension sets $(\bar{L},\bar{L},2\bar{L})$. Finally, the final snapshots of the NVT simulations are taken as the starting structures of new 60~ns NVT simulations for which the cell dimensions are  $(\bar{L},\bar{L},6\bar{L})$. The simulated systems present thus two liquid/vapor interfaces. 

\begin{figure}[H]
\includegraphics[scale=.7]{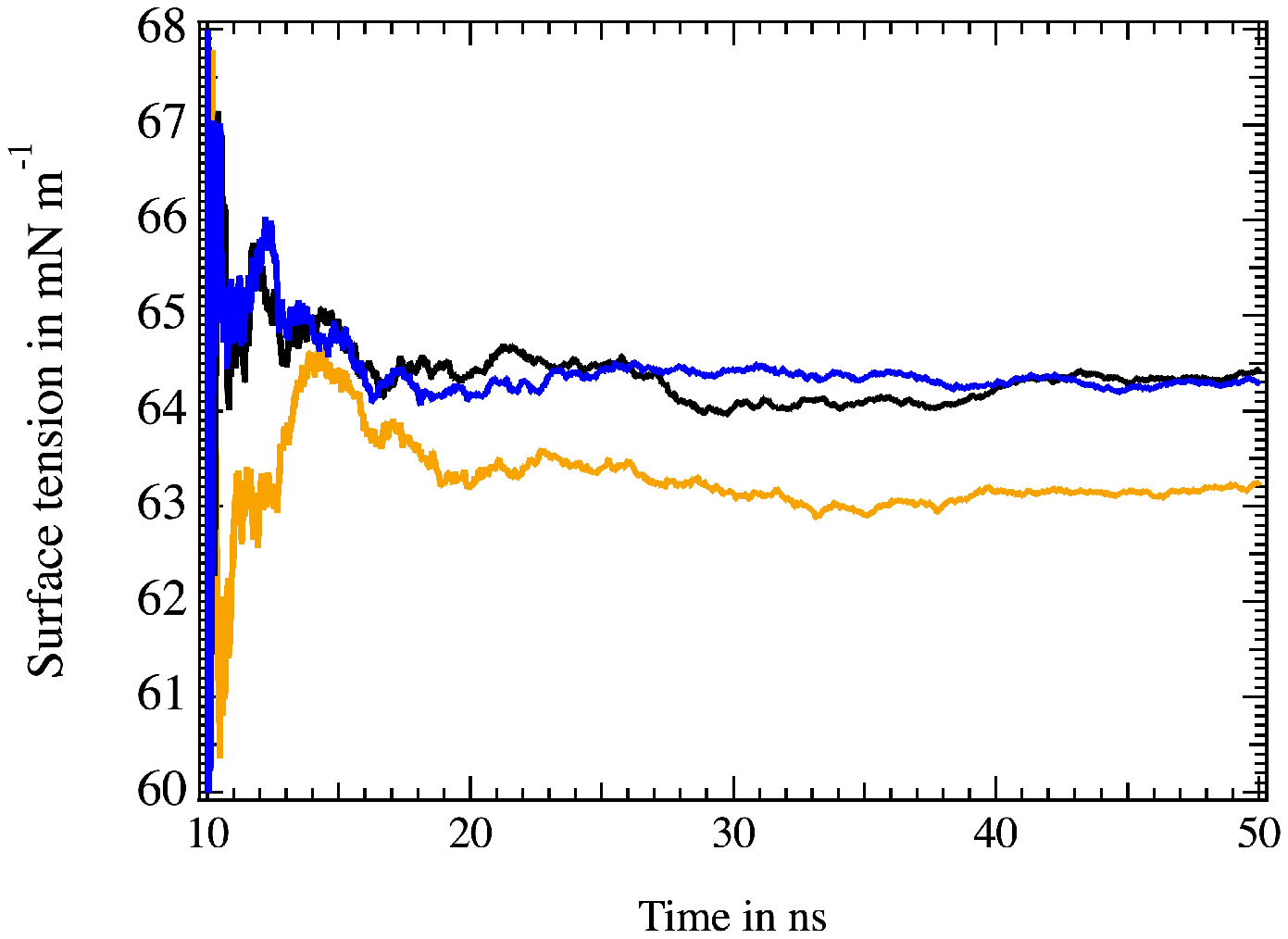}
\caption{Temporal evolutions of the surface tensions of planar interfaces. Black : 0.6$m$ NaCl solution; blue : 0.2$m$ NaCl solution; orange : neat water.}
 \label{fig:surface-tension}
\end{figure}

Along the 60~ns NVT simulations and once an initial relaxation phase of 10~ns is achieved, we computed the surface tensions $\gamma(\mathrm{T})$ of liquid water and of 0.2 and 0.6$m$ NaCl aqueous solutions at a temperature T = 300 K by means of the test-area simulation method \cite {gloor05}

\begin{equation}
\gamma (\mathrm{T}) = \left( \dfrac{\Delta A}{\Delta S} \right)_{N,V,T},
\end{equation}
here, $\Delta A$ is the change in the free energy for an infinitesimal change in the interfacial area $S$. The $\gamma (\mathrm{T})$'s are computed according to a Free Energy Perturbation scheme (the sampling interval is 50 fs), with the two perturbed states corresponding to the altered surface area $S_{\pm} = L_xL_y (1 \pm \delta_s)$. $x$ and $y$ are the two dimensions orthogonal to the $z$-axis (the perturbation is equally applied to both). We set $\delta_s$ to  5.10$^{-3}$ as recommend in Ref. \cite {vega07}. The temporal evolutions of the surface tensions for our planar interfaces are plotted in Figure \ref{fig:surface-tension}.

%Introduction

\begin{figure}[H]
\includegraphics[scale=.7]{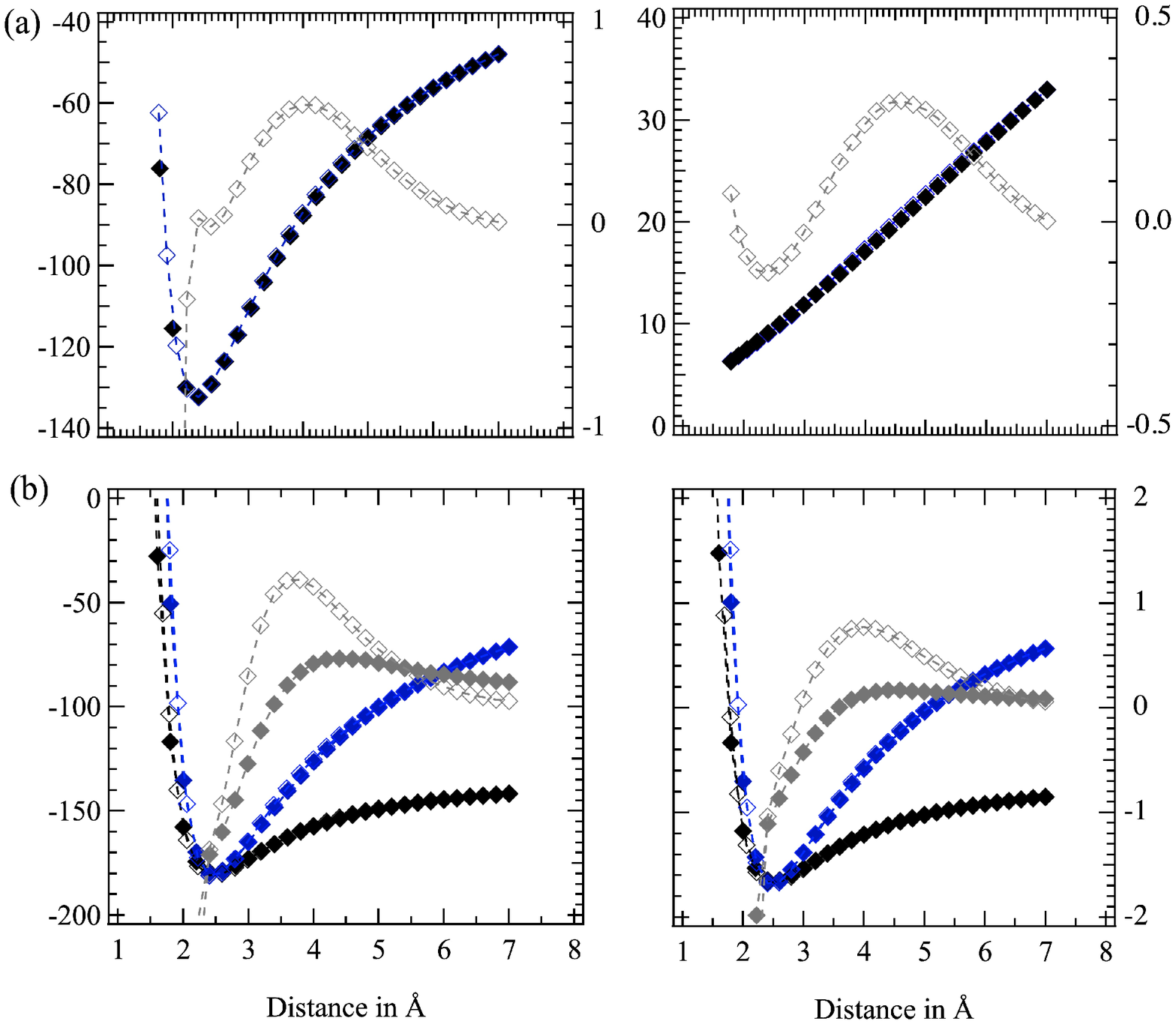}
 \caption{ (a) \ce{[Na^+,Cl^-]} ion pair interaction energy (in \kcal, plot on the left) and total dipole moment (in Debye, plot on the right) in gas phase as a function of the inter ionic distance  from \emph{ab initio} MP2/CBS computations (full black symbols) and from our force field (empty blue symbols). Dashed grey lines and symbols: differences between force field and \emph{ab initio} data (right axes). (b) Interaction energies (in \kcal)  of the  ionic trimers \ce{[(Na^+)_2,Cl^-]}  (left) and \ce{[Na^+,(Cl^-)_2]} (right). In these trimers, the single ion is set in between the two ions of opposite charge. Full and empty symbols: quantum and force-field data, respectively. Blue symbols (left axis): trimer interaction energies corresponding to symmetric dissociation of the two charge-like ions. Black symbols (left axis): trimer interaction energies corresponding to the dissociation energy of one of the two charge-like ions. Grey lines and symbols (right axes): force field/quantum differences in the trimer energies.}  \label{fig:ion_pair_gas_phase}
 \end{figure}
\newpage

\begin{figure}[H]
\includegraphics[scale=1]{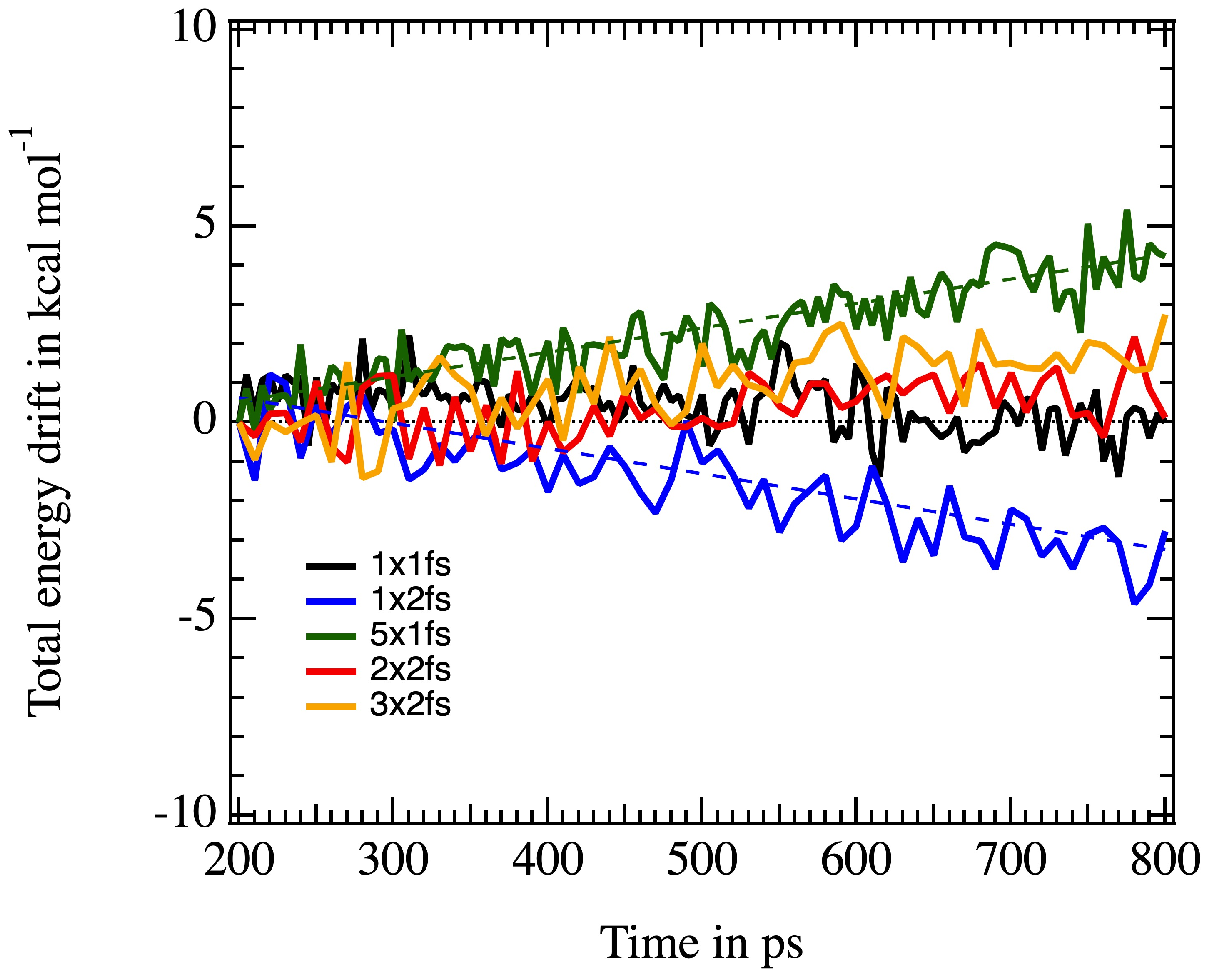}
 \caption{Total energy drifts along 1~ns scale MD simulations in the NVT ensemble at ambient conditions of a cubic box comprising 10 NaCl pairs and about 1000 water molecules. That box is simulated using Particle Mesh Ewald summation techniques, periodic conditions and the polarizable force fields according to the MD protocol detailed in the main manuscript. Here we used five different sets of (X$\times$Y) time steps (in fs) to update the short and long range forces as solving the equations of motion using the r-RESPAp multiple time step scheme, from (1$\times$1 fs) up to (2$\times$6~fs). The temporal evolutions of the energy drift are here all linear on average, the largest slope is about 7 $10^{-3}$ \kcal~ps$^{-1}$ in absolute values. }
  \label{fig:energy_drift}
 \end{figure}
\newpage

\begin{figure}[H]
\includegraphics[scale=.75]{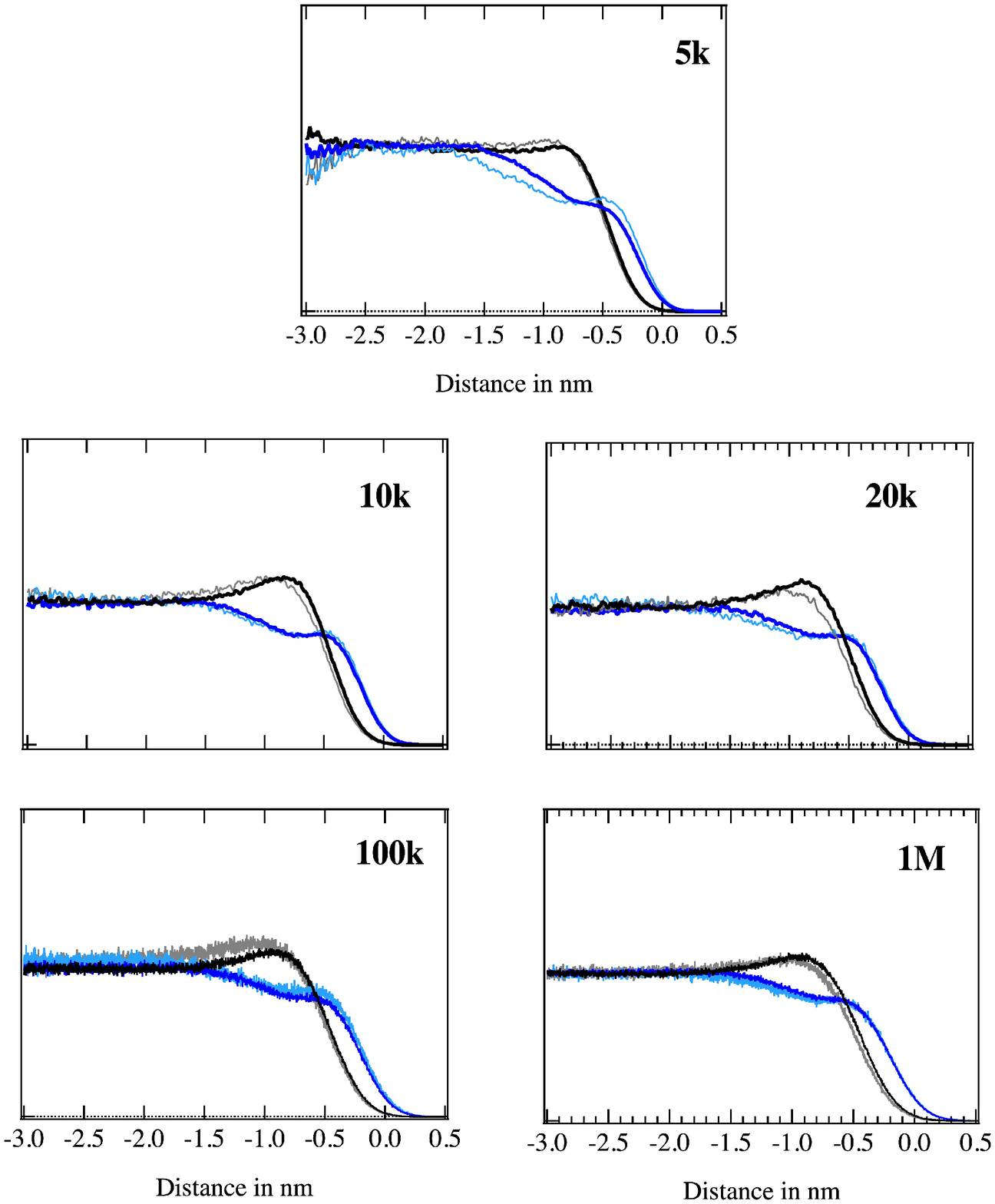}
 \caption{Normalized distributions of ions as a function of their distance from the droplet boundary. From up to down 5k, 10k, 20k, 100k and 1M droplets. Black  : \chem{Na^+} distributions, blue : \chem{Cl^-} distributions. Light colors correspond to 0.2$m$ droplets.}
  \label{fig:ion_distribution}
 \end{figure}
\newpage

\begin{figure}[H]
\includegraphics[scale=.8]{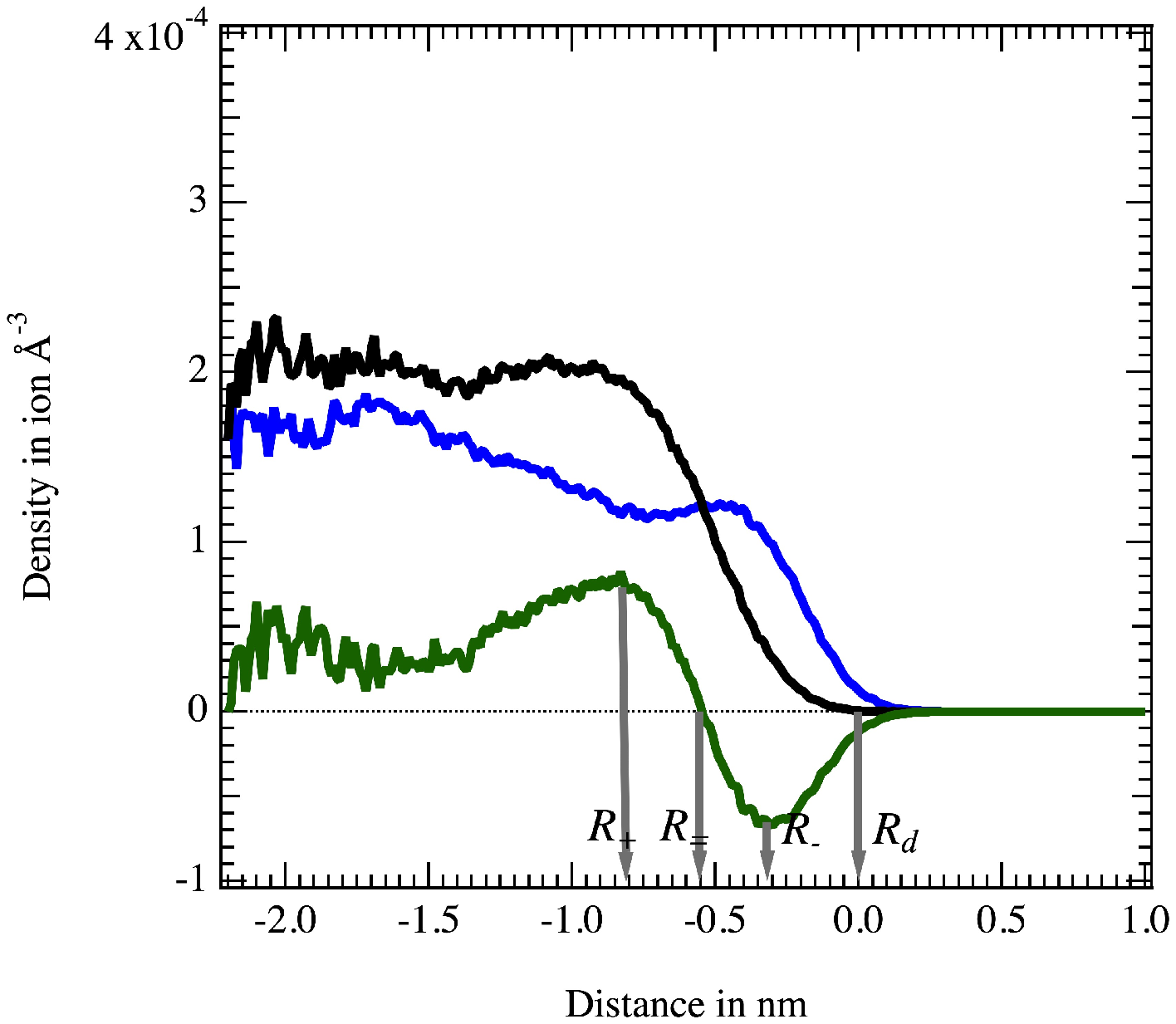}
 \caption{Ion densities as a function of the ion distance from the equi molar radius $R_d$ for the 2k 0.2$m$ droplet. \ce{Na^+} and \ce{Cl^-} densities are plotted in black and blue lines, respectively. Green line: difference $\Delta \rho^i(r)$ in the \ce{Na^+} and \ce{Cl^-} densities. The vertical grey arrows are located at the minimum, zero and maximum values of the $\Delta \rho^i(r)$ function and at the mean droplet radius $R_d$.}
  \label{fig:densities_2k}
 \end{figure}
\newpage

\begin{figure}[H]
\includegraphics[scale=.7]{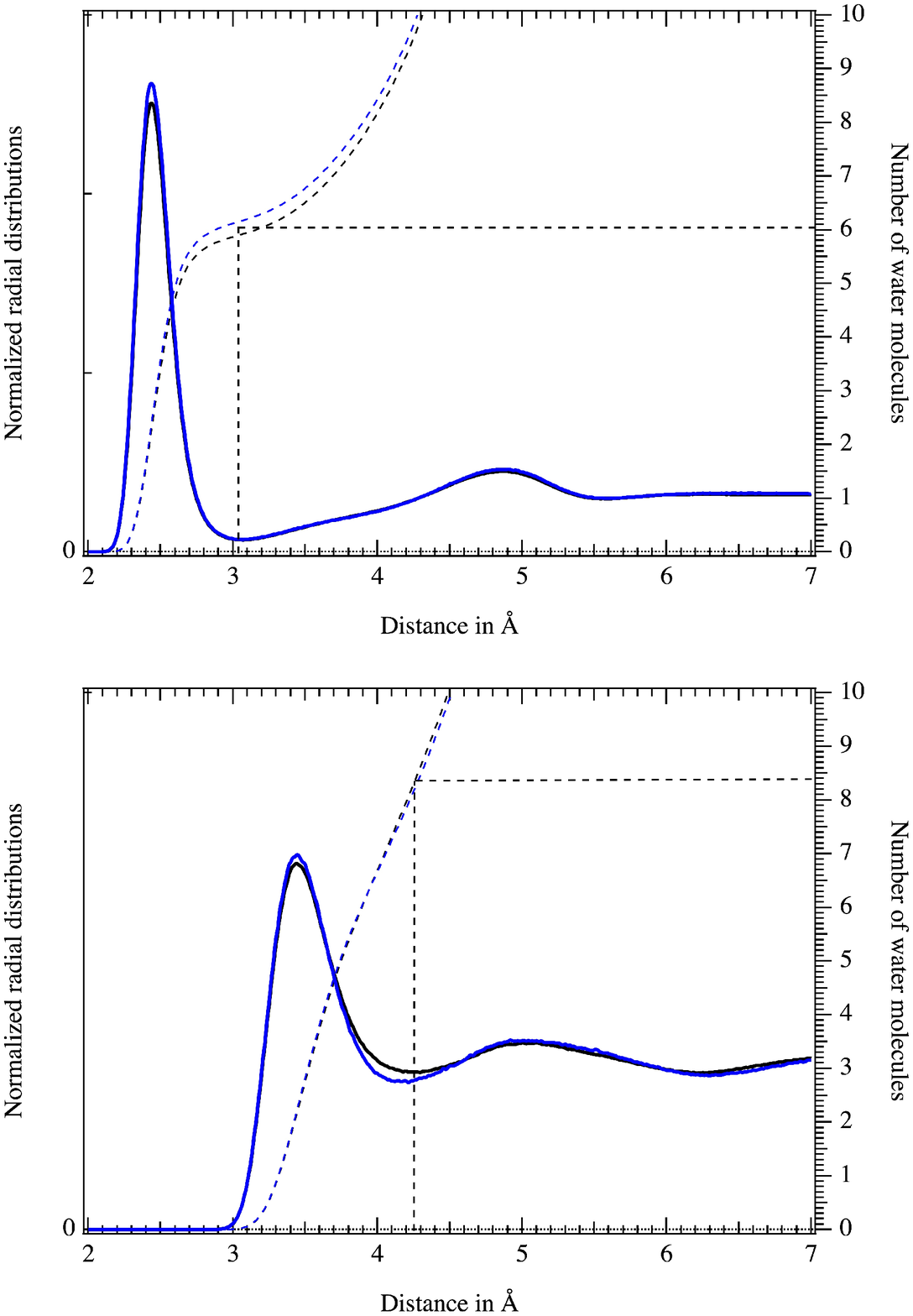}
 \caption{Ion water radial distribution functions (left axes) and their integrals (right axes) as a function of the ion/oxygen distance for the 0.2$m$ (blue) and 0.6$m$ (black) 10k droplets. Up and down : \na and \cl functions, respectively. The uncertainty regarding the position of the second minimum of the distribution functions is about 0.1 \AA. That yields an uncertainty regarding the number of water molecules in the ion first coordination sphere that amounts to 0.2 (\na) and 0.5 (\cl).}
  \label{fig:ion_water_distribution}
 \end{figure}
\newpage

\begin{figure}[H]
\includegraphics[scale=.6]{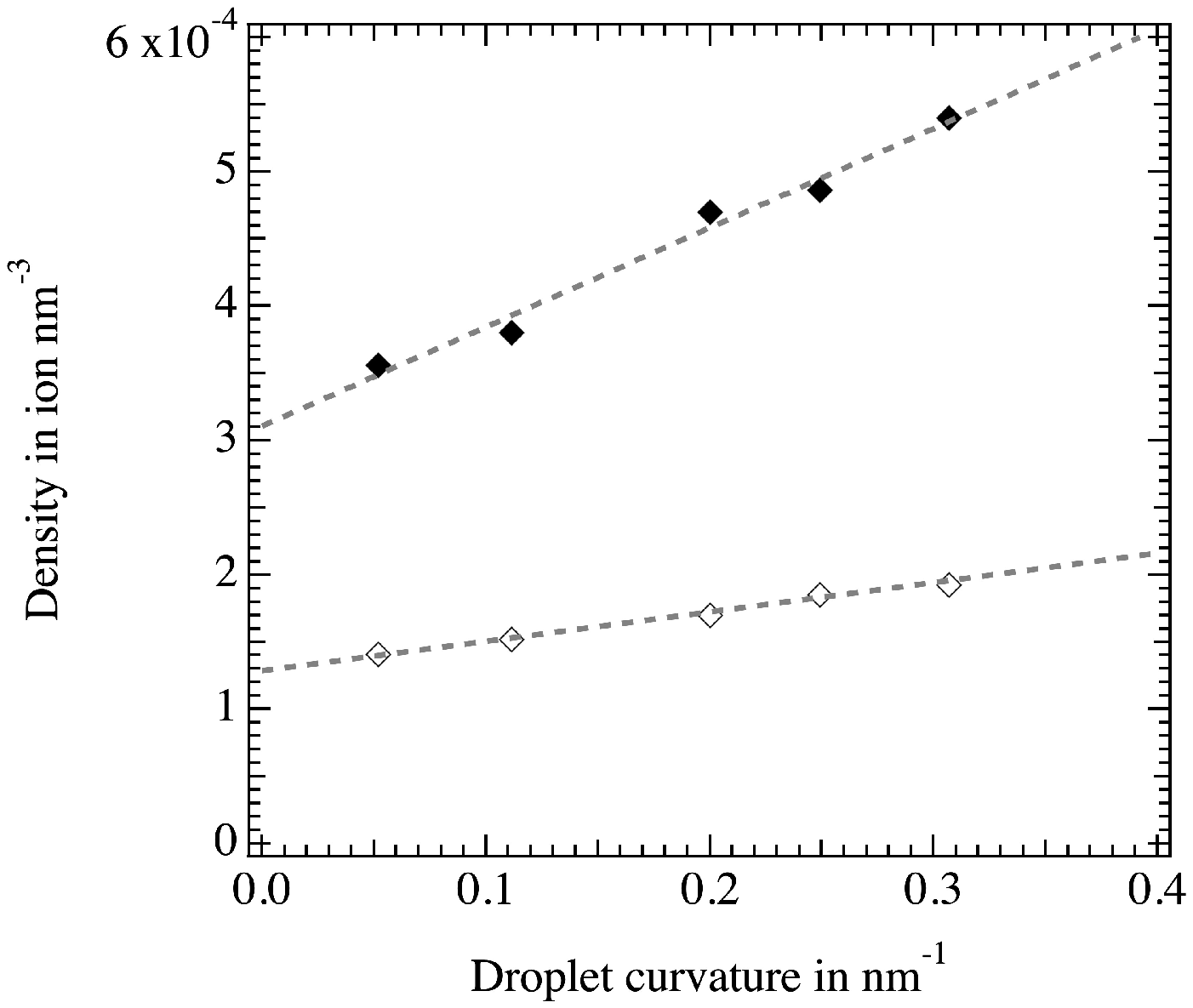}
\caption{Mean densities $\rho_d^{\ce{Na^+}}$ within the droplet bulk-like core domains as a function of the inverse of the droplet radius $R_d$. Empty and full symbols: 0.2 and 0.6$m$ data, respectively. In dashed line the linear $R_d^{-1}$ functions that best reproduce the densities (the regression coefficients are > 0.98).}
  \label{fig:ion_densities}
 \end{figure}
\newpage

\begin{figure}[H]
\includegraphics[scale=.8]{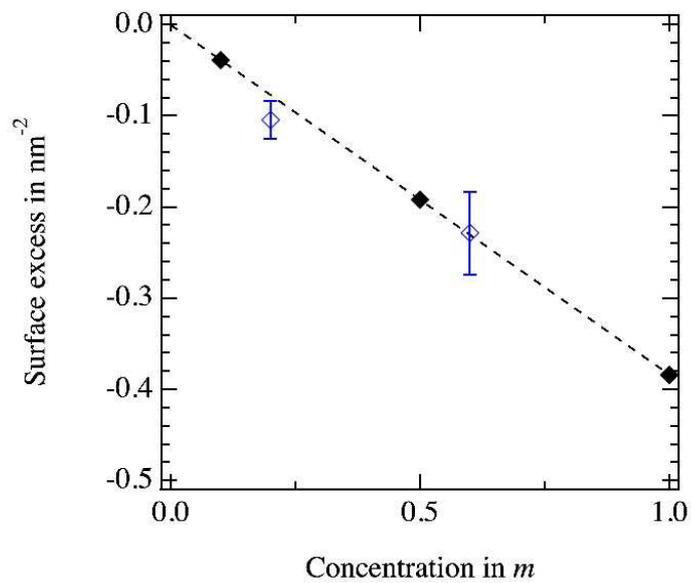}
 \caption{Total ion surface excess values $\Gamma_\mathrm{tot}$ as a function of the droplet curvature $R_e^{-1}$. Blue and black: 0.2 and 0.6$m$ data, respectively; dashed lines:  linear regression fits of data corresponding to 10k-1M droplets.}
  \label{fig:surface_excess}
 \end{figure}
\newpage

\begin{figure}[H]
\includegraphics[scale=.8]{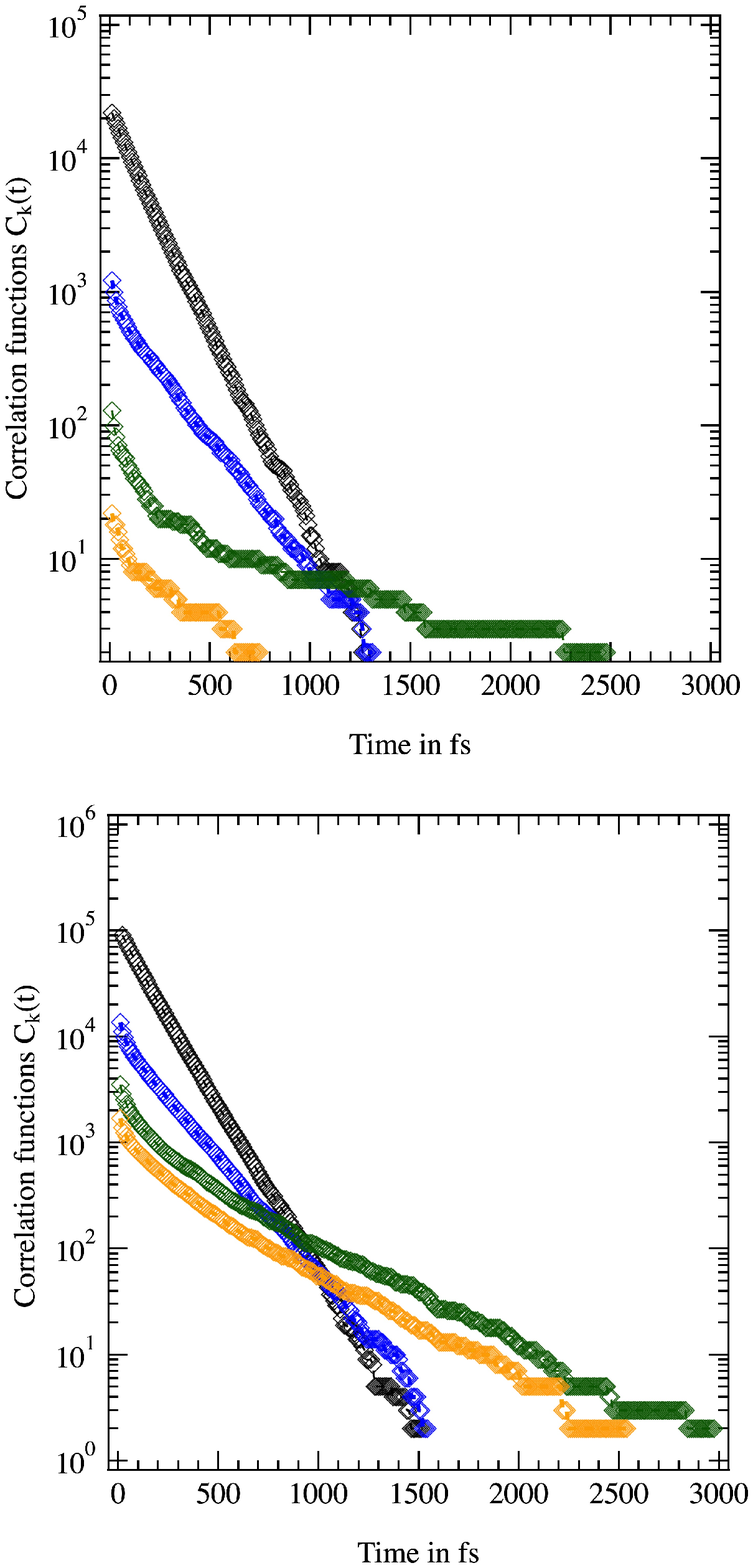}
\caption{Cluster temporal correlation functions $C_k(t)$. Black, blue, green and orange : data for 2, 3, 4 and 5-sized ion clusters. Up and down : 0.2 and 0.6$m$ data. The linear regression to compute the time $\tau^*$ discussed in the main manuscript are performed on the initial segment of the functions.}
  \label{fig:correlation_ck}
 \end{figure}
\newpage

\begin{figure}[H]
\includegraphics[scale=.8]{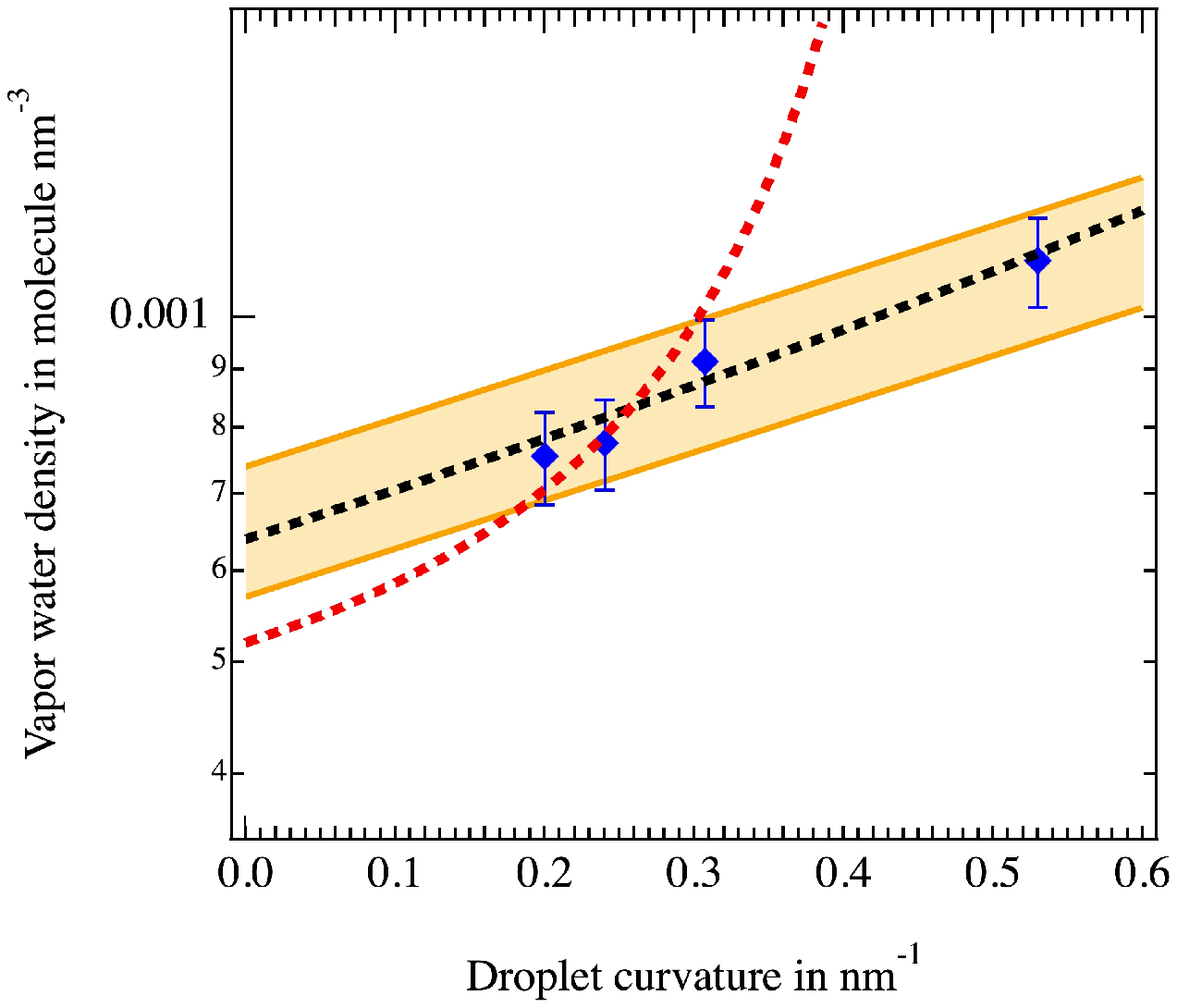}
 \caption{Water vapor density (ln scale) as a function of the droplet curvature $R_d^{-1}$ for pure aqueous droplets whose molecular size ranges from 1k to 20k. Bold dashed line : interpolation of the raw density data according to the Kelvin relation, using a Tolman length  value $\delta_\gamma$ = -0.1 nm and a planar interface extrapolated value $\rho_\infty^g$ set to  0.64 10$^{-3}$ molecule nm$^{-3}$. Oranges bold lines : best standard Kelvin terms ($\delta_\gamma$ = 0) allowing to reproduce the densities to which we add/subtract the density uncertainties. Red bold lines : best interpolation of salty droplet data corresponding to a Tolman length of -0.85 nm, see main manuscript.}
  \label{fig:water_surface_tension}
 \end{figure}
\newpage

\begin{figure}[H]
\includegraphics[scale=.8]{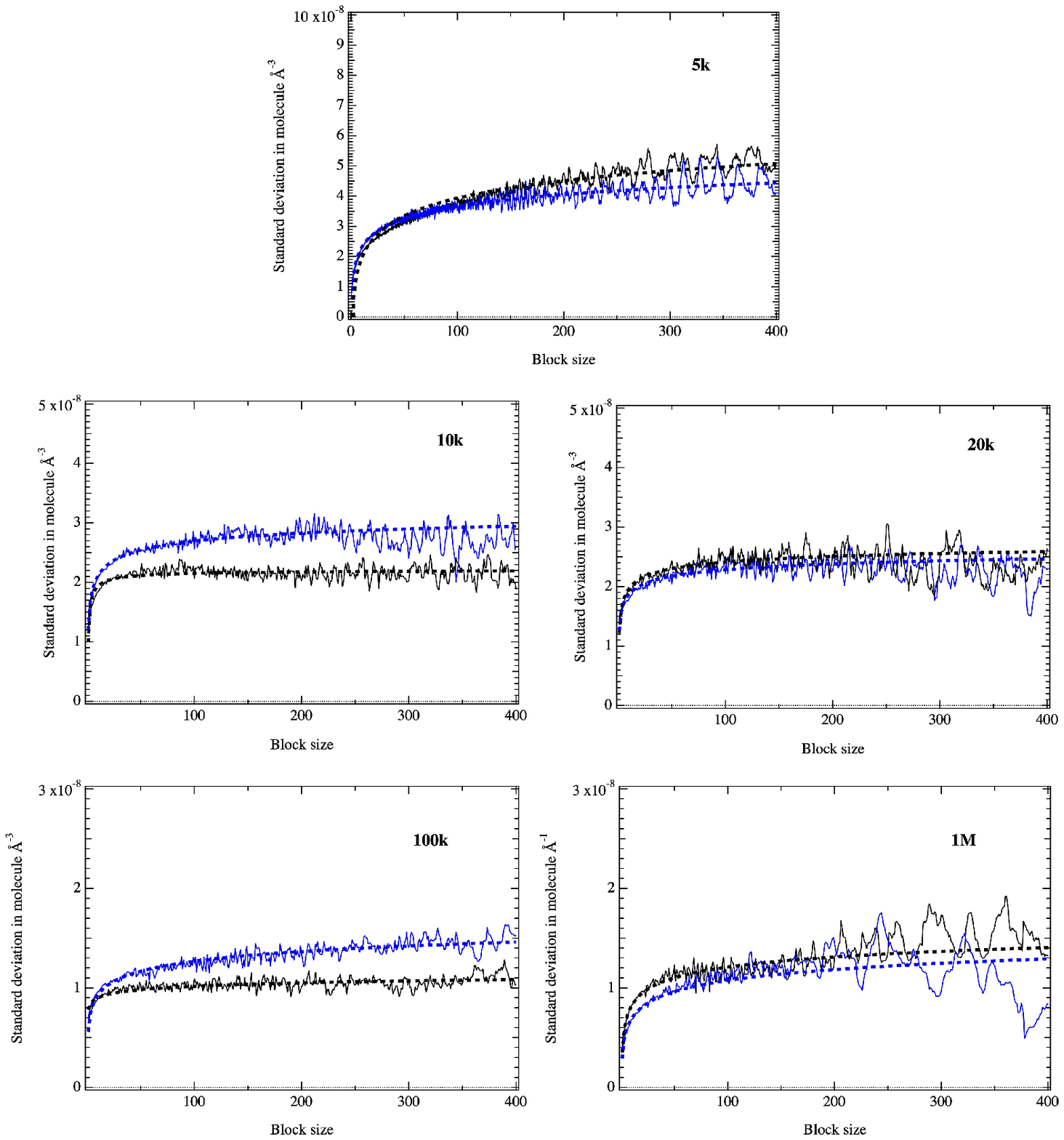}
 \caption{Standard block analysis of the water vapor densities. The density standard deviation values are plotted as a function of the block size $b$ of contiguous temporal data. From up to down : 5k to 1M droplets. Full back and blue lines : 0.2 and 0.6$m$ data. Dashed lines : functions $\alpha + \beta /b^\gamma$ whose parameters $\alpha$, $\beta$ and $\gamma$ are adjusted to best reproduce the raw uncertainty data. The upper values of the adjusted parameters $\alpha$ for each kind of 0.2/0.6$m$ droplets are the uncertainties of the mean water vapor density values discussed in the main manuscript, regardless of the salt concentration.}
  \label{fig:block_analysis}
 \end{figure}
\newpage

\begin{figure}[H]
\includegraphics[scale=.6]{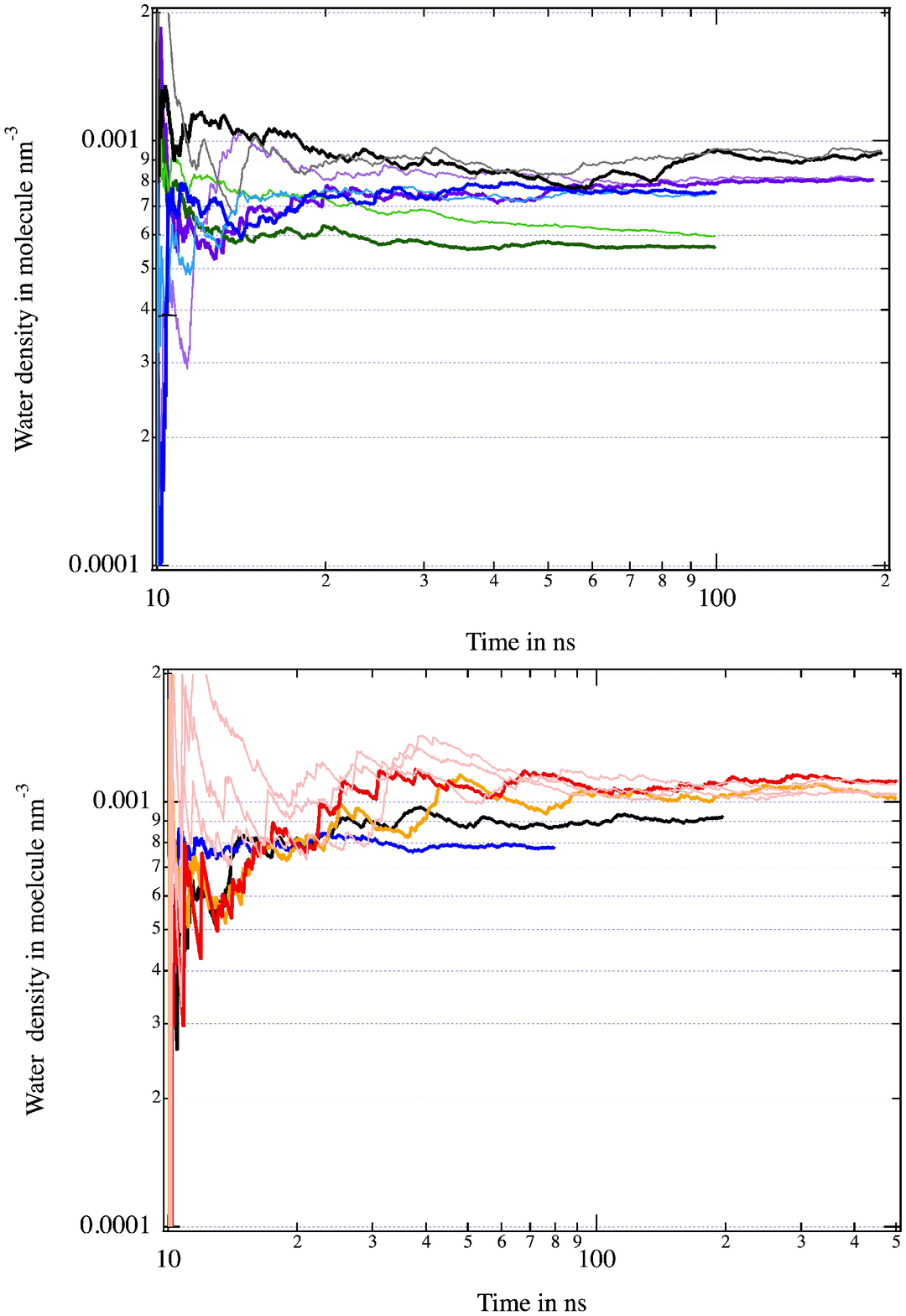}
 \caption{Up : temporal evolutions of the water vapor density for the 0.2$m$ (thin lines) and 0.6$m$ (bold lines) 5k (black), 10k (violet), 20k (blue) and 100k (green) salty droplets. Down : temporal evolutions of the water vapor density for 1k (red bold and thin lines, each line corresponds to an independent 500~ns simulation), 5k (orange), 10k (black) and 20k (blue) pure aqueous droplets.}
  \label{fig:temporal_vapor}
 \end{figure}
\newpage

\begin{figure}[H]
\begin{subfigure}{1.\linewidth}
\centering
\includegraphics[scale=.65]{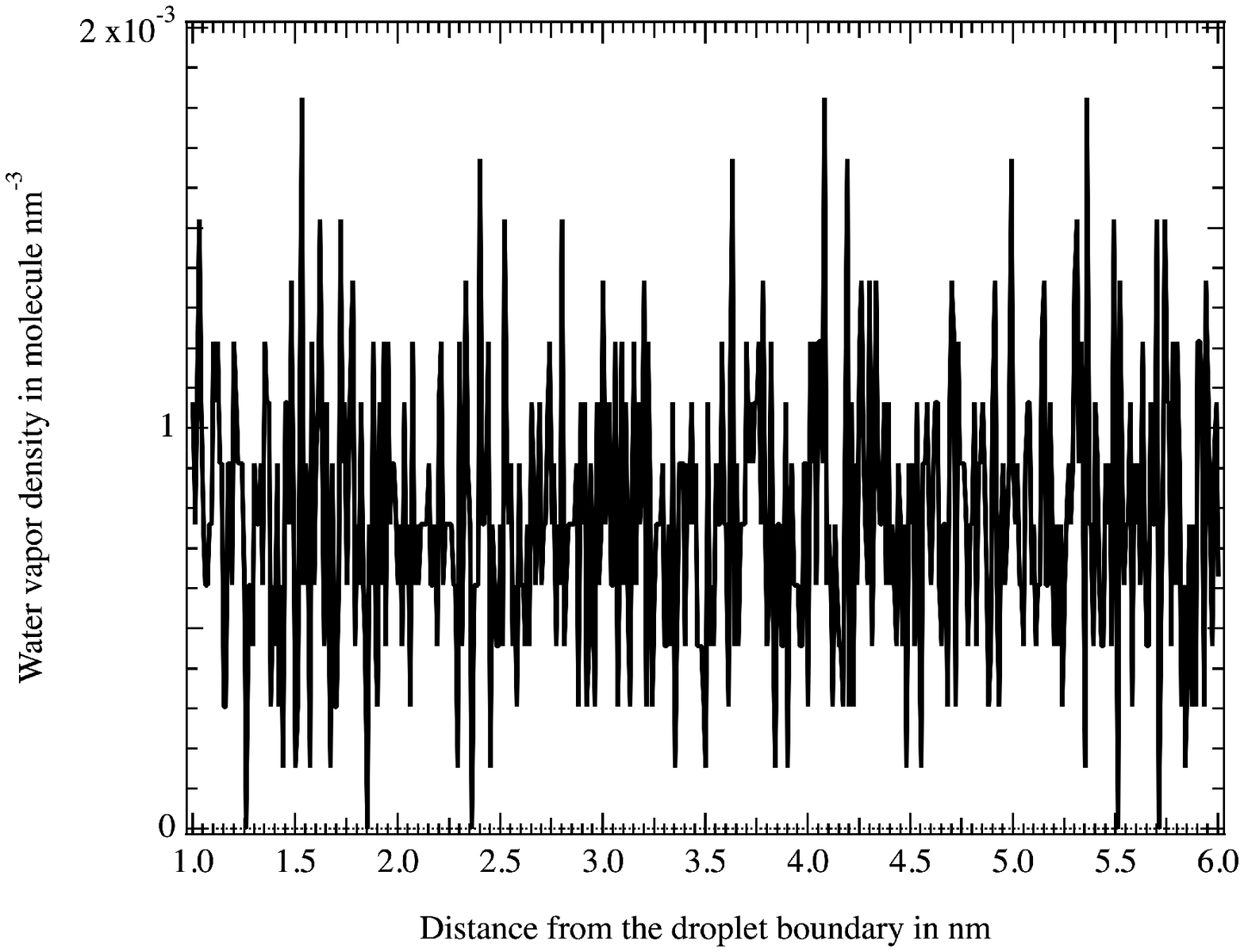}
\end{subfigure}
 \begin{subfigure}{1.\linewidth}
 \centering
\includegraphics[scale=.65]{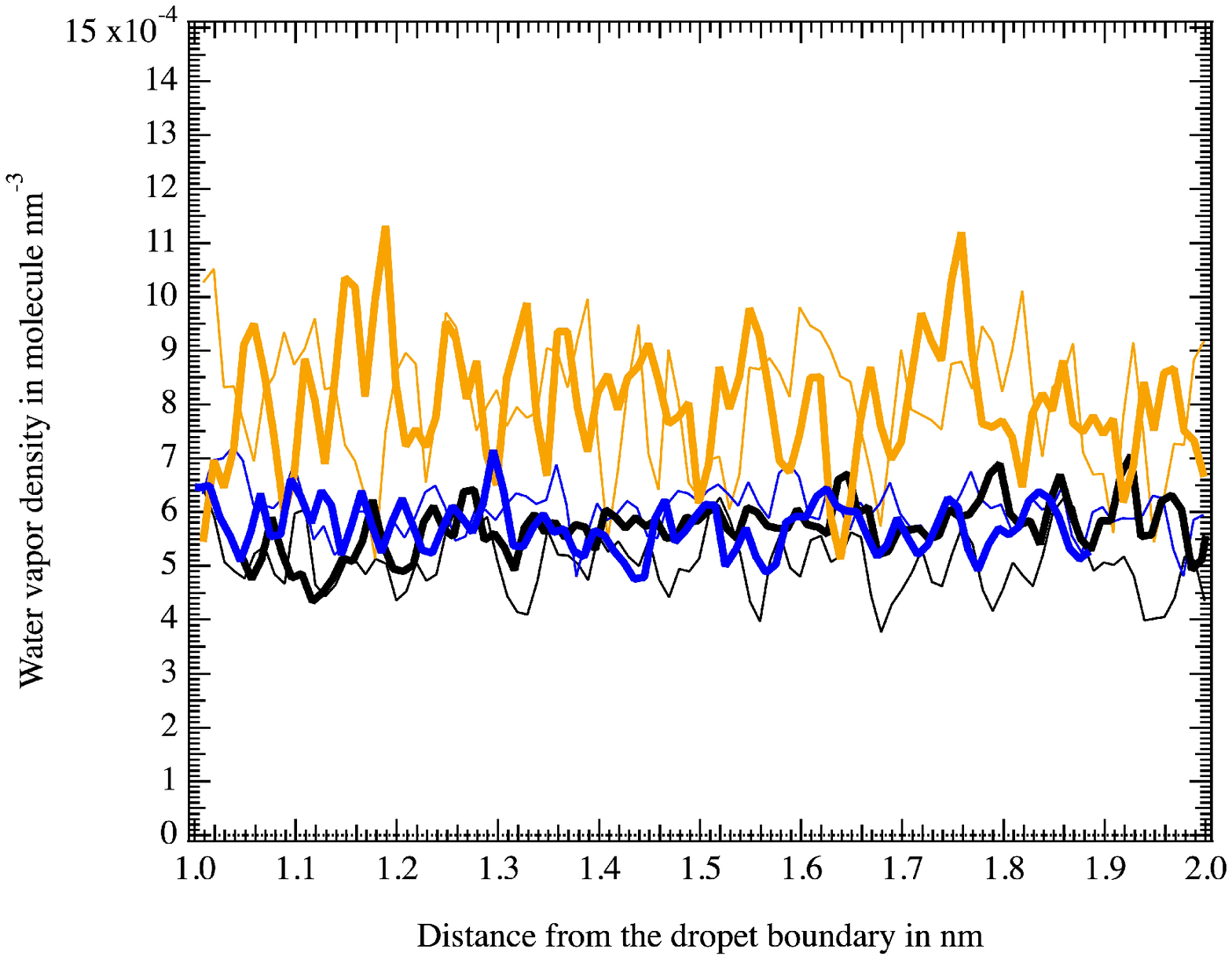}
\end{subfigure} 
  \caption{Up : raw mean water vapor density as a function of the distance from the liquid/gas phase planar interface for the 0.2$m$ NaCl aqueous solution as computed along a 100~ns simulation. The vapor density profile for the 0.6$m$ NaCl solution is similar. Down : raw mean water vapor density as a function of the distance from the spherical interfaces of 0.2$m$ (thin lines) and 0.6$m$ (bold lines) NaCl aqueous droplets whose molecular sizes are 10k (orange), 100k (blue) and 1M (black) as computed along our simulations. The densities are averaged on rectangular or spherical shells whose width is set to 0.01 nm.}
   \label{fig:vapor_density}
 \end{figure}
\newpage

\section*{Tolman length and cumulative number of cloud condensation nuclei}

The SSA supersaturation $s$ (in percent) is defined from the K\"ohler relation according to  

\begin{equation}
s = 100 \times \left [1 - \dfrac{V_{wet} - V_{dry}}{V_{wet} + (1 - \kappa) V_{dry}} \times \exp \left(  \dfrac{2\gamma_\infty}{\mathrm{RT} \rho_\infty (R_d + 2\delta_\gamma)} \right) \right].
\end{equation}

\begin{figure}[H]
\begin{subfigure}{1.\linewidth}
\centering
\includegraphics[scale=.8]{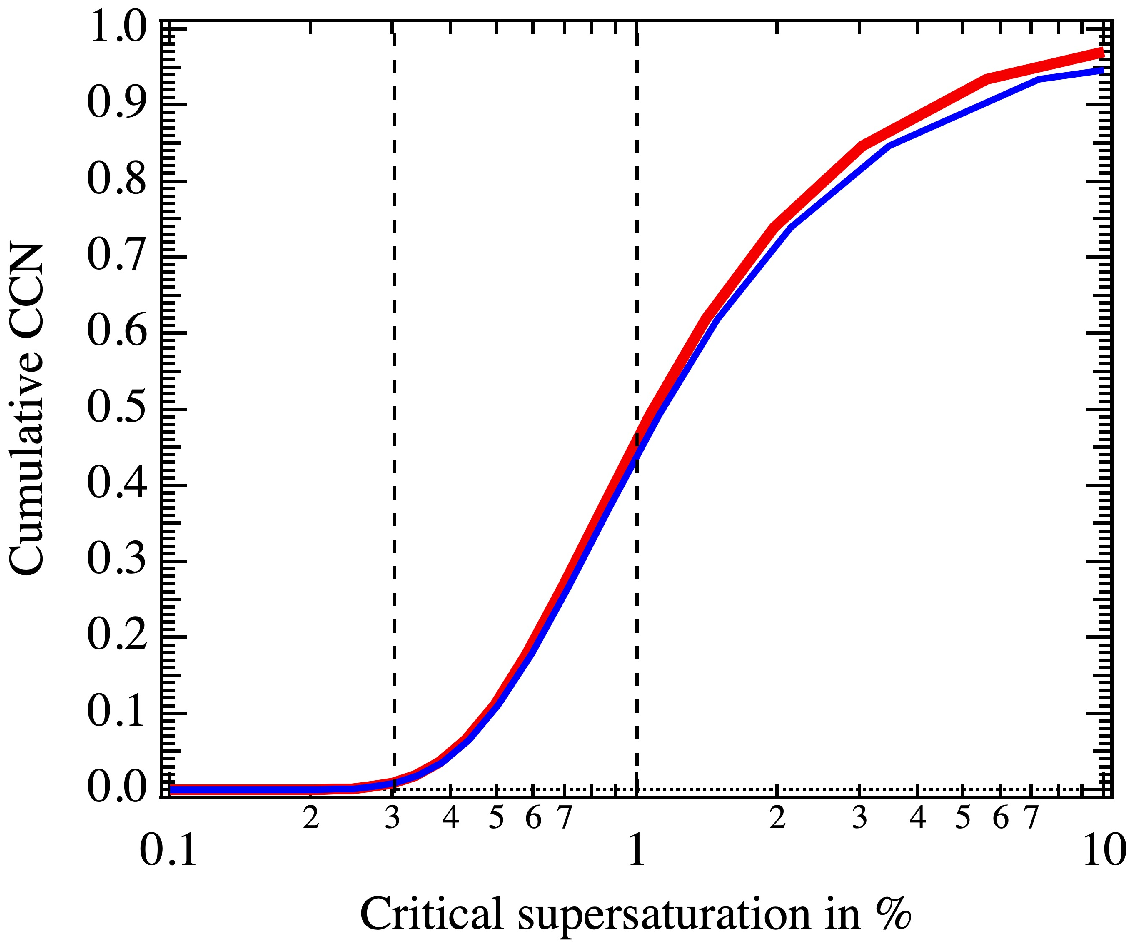}
\end{subfigure}
 \begin{subfigure}{1.\linewidth}
 \centering
\includegraphics[scale=.8]{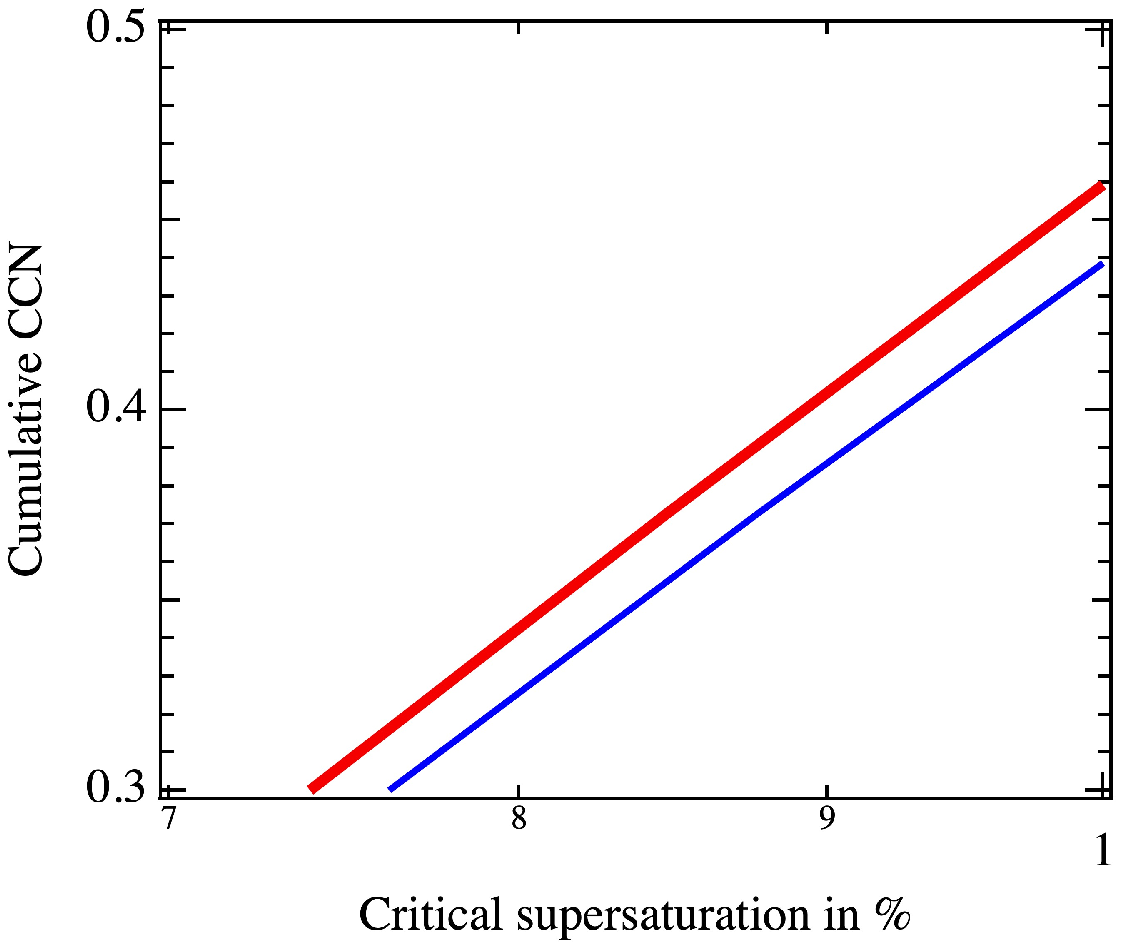}
\end{subfigure} 
 \caption{Up : cumulative numbers of CCN for a normalized gaussian distribution of sub-micron NaCl aqueous SSAs as a function of the SSA $s_{crit}$ value for a Tolman length of 0 (red) and of -1 nm (blue). Down : details of the above plots.}
  \label{fig:nuclei}
 \end{figure}

$V_{wet} - V_{dry}$ measures the difference in the volumes of a SSA in equilibrium with a water vapor phase (the SSA radius is then $R_d$) and in its dry state. $\kappa$ is the SSA hygroscopicity parameter. We set $\kappa$ to its standard value for NaCl aqueous droplets ($\kappa = 0.67$). For our purpose we set here $\gamma_\infty$ to the mean value computed for the planar interfaces corresponding to 0.2 and 0.6$m$ NaCl aqueous solutions and we set the water liquid density $\rho_\infty$ to its value for neat water at ambient conditions (33.3 molecule nm$^{-3}$). In Figure \ref{fig:nuclei} we plot the cumulative numbers of cloud condensation nuclei as a function of the critical supersaturation value $s_\mathrm{crit}$ (\emph{i.e.} the largest value of $s$ for a SSA corresponding to a dry volume $V_{dry}$ ) for a Tolman length set to 0 and of -1 nm. See ref. \cite{farmer15} for details about that kind of plots. For that figure we consider a size (diameter) distribution of SSAs (in their dry state)  corresponding to a normalized gaussian centered at 20 nm and which vanishes at 100 nm. 

\bibliography{bibliography}

% End of the document